\documentclass[11pt]{article}
\usepackage{multicol}
\usepackage{graphicx}
\usepackage{helvet}

\newcounter{myFCounter}[section]
\usepackage{graphicx}

\usepackage{amssymb}
\usepackage{graphicx}
\usepackage{color}
\usepackage{amscd}

\topmargin by -\headsep \textheight 9in \oddsidemargin 0pt
\evensidemargin \oddsidemargin \marginparwidth 0.5in \textwidth
6.5in
 \catcode`\@=11
\def\marginnote#1{}
\begin{document}
\title{The use of oscillatory signals in the study of genetic networks }
\author{Ovidiu Lipan \thanks{email: olipan@mail.mcg.edu}\\
\vspace{-0.5cm}
     \tiny{\it{Center for Biotechnology and Genomic Medicine}}\\
     \vspace{-0.5cm}
     \tiny{\it {Medical College of Georgia}}\\
     \vspace{-0.5cm}
      \tiny{\it {1120 15th St.,CA-4124}}\\
     \tiny{\it{Augusta, Georgia 30912, USA}} \\
    \and
    Wing H. Wong \thanks{email:  wwong@stat.harvard.edu}\\
    \vspace{-0.5cm}
   \tiny{\it {Departments of Statistics and Biostatistics}}\\
   \vspace{-0.5cm}
   \tiny{ \it {Harvard University}}\\
   \vspace{-0.5cm}
     \tiny{\it {1 Oxford Street}}\\
     \vspace{-0.5cm}
     \tiny{\it {Cambridge,Massachusetts 02138, USA}}}
     \date{Submitted to PNAS on May 27th 2004. The paper is under
     consideration.}
\maketitle
\begin{abstract}
\bf{The structure of a genetic network is uncovered by studying
its response to external stimuli (input signals). We present a
theory of propagation of an input signal through a linear
stochastic genetic network. It is found that there are important
advantages in using oscillatory signals over step or impulse
signals, and that the system may enter into a pure fluctuation
resonance for a specific input frequency.}
\end{abstract}

\parskip 1pt
%

%\begin{multicols}{2}
The nature of a physical system is revealed through its response
to external stimulation. The stimulus is imposed upon the system
and its effects are then measured, Fig.1(a). This approach is
widely used in biology: a cell culture perturbed with a growth
factor, a heat shock etc. The data measured contain the initial
information encoded into the stimulus plus the information about
the intrinsic characteristics of the system. The more parameters
the experimentalist can adjust to craft the perturbation stimulus,
the more information about the system can be revealed. In recent
years we witnessed a tremendous increase in measurement
capabilities (e.g., microarray and proteomic technologies, better
reporter genes). However, the success of the systems approach to
molecular biology depends not only on the measurement instruments,
but also on an effective design and implementation of the input
stimulus, which has not been thoroughly explored. Traditionally,
two types of time dependent stimuli are at work in molecular
biological experiments \cite{Gardner}, \cite{Vance}. For example a
step stimulus is obtained when at one instant of time a growth
factor is added to the medium, graph (a) in Fig.1(a). The stimulus
from Fig.1(a) graph (b), is a superposition of two step stimuli.
The investigator can control the height of the step stimulus (the
concentration of the growth factor) or the time extension of the
heat shock. The cells respond to these stimuli only transiently.
The response is dampened after some time and becomes harder to
detect it from noise. To overcome the noise, the concentration of
the stimulus is typically increased to the point where the
strength of the stimulus raises far above its physiological range.

 We propose to implement a molecular switch at the level of gene
promoter and use it to impose an oscillatory stimulus. In the
absence of experimental noise, any stimulus can be used to
determine the genetic network input-output properties. However, in
the presence of experimental noise, the oscillatory input has many
advantages: (1) the measurements can be extended to encompass many
periods so the signal-to-noise ratio can be dramatically improved;
(2) the measurement can start after transient effects subside, so
that the data becomes easier to be incorporated into a coherent
physical model; (3) an oscillatory stimulus has more parameters
(period, intensity, slopes of the increasing and decreasing
regimes of the stimulus) than a step stimulus. As a consequence,
the  measured response will contain much more quantitative
information. Experimental results from neuroscience prove that
oscillatory stimulus can modulate the mRNA expression level of
genes. For example, c-fos transcription level in cultured neurons
is enhanced $400\%$  by an electrical stimulus at $2.5$ Hz and
reduced by $50\%$at $0.01$ Hz, \cite{Seo}. Also, the mRNA levels
of cell recognition molecule L1 in cultured mouse dorsal root
ganglion neurons change if the frequency of the electric pulses is
varied. The expression level of L1 decreases significantly after 5
days of $0.1$ Hz stimulation but not after 5 days of 1 Hz
stimulation \cite{Itoh}. To extend the oscillatory approach to
other type of cells, a two-hybrid assay, \cite{Quail}, can be used
to implement a molecular periodic signal generator, Fig.1(c). The
light-switch is based on a molecule (phytochrome in \cite{Quail})
that is synthesized in darkness in the Q1 form. When Q1 form
absorbs a red light  photon ( wavelength 664 nm) it is transformed
into the form Q2. When Q2 absorbs a far red light ( wavelength 748
nm) the molecule Q goes back to its original form, Q1. These
transitions take milliseconds. The protein P interacts only with
the Q2 form, recruiting thus the activation domain (AD) to the
target promoter. In this position, the promoter is open and the
gene is transcribed. After the desired time elapsed, the gene can
be turned off by a photon from a far red light source. Using a
sequence of red and far red light pulses the molecular switch can
be periodically opened and closed.

\begin{figure}
\centering
\includegraphics[width=14cm]{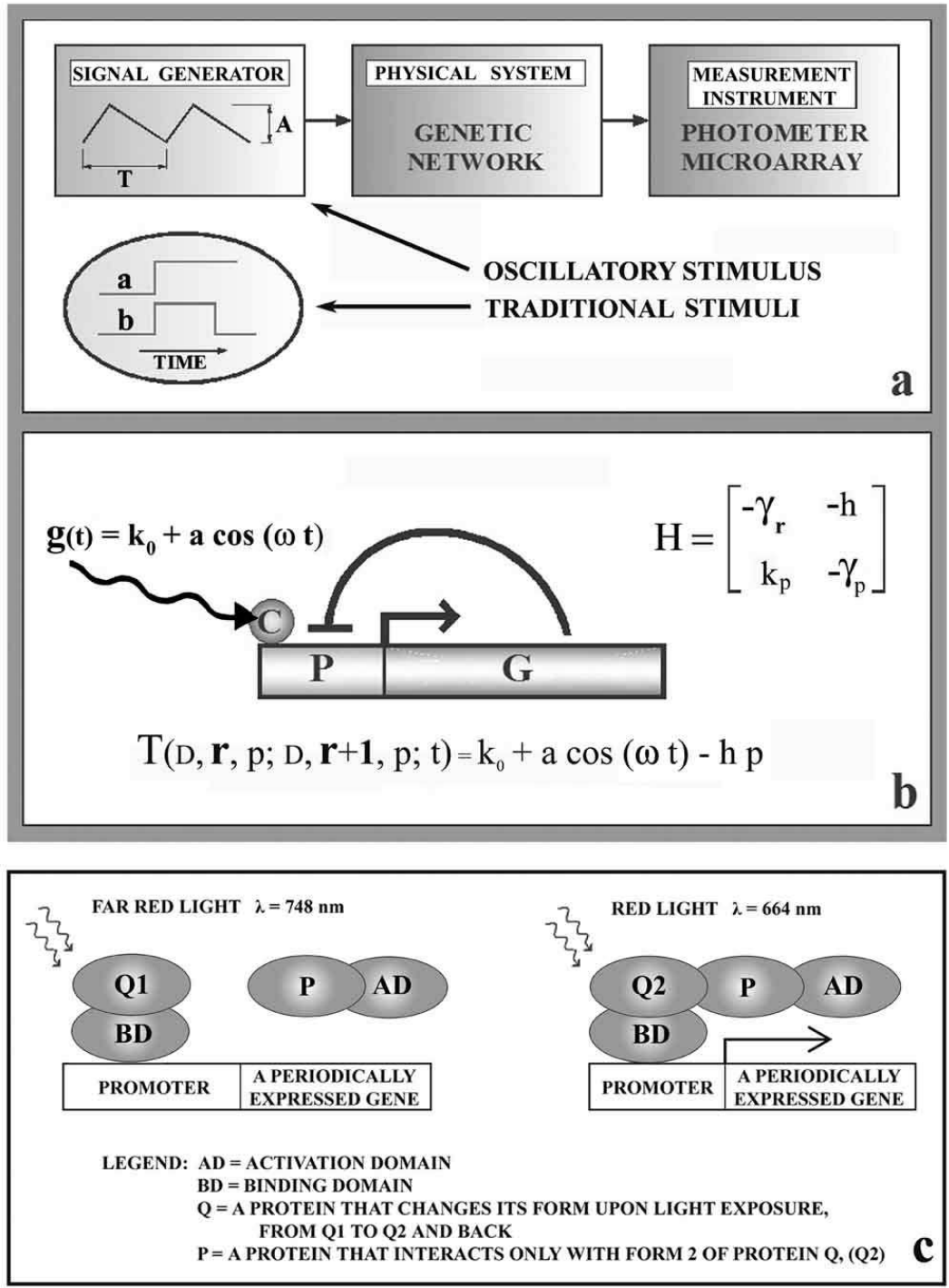}
 \caption{(a): The
genetic network response depends on the type of the applied
stimulus. (b): An autoregulatory network. The gene G is under the
influence of a cofactor C that rhythmically modulates the activity
of the promoter P. The matrix H contains the parameters that
dictates the transition probabilities of the stochastic model. The
transition probability per unit time from $r$ to $r+1$ mRNA
molecules, $T(D,r,p;D,r+1,p;t)$, is modulated by the oscillatory
signal generator. The DNA, $\it D$, and the protein, $\it p$, do
not change in this transition. (c), adapted from [5]. The gene is
turned on with a red light pulse of wavelength $\lambda =664 $ nm.
With a far red light of wavelength $\lambda =748$ nm the gene is
turned off.}
\end{figure}

 There are four input parameters that can be varied:
the period (T), the time separation between the pulses (s), and
the amplitude (A) of the red and the far red pulses. The mRNA
concentration profile will depend on these parameters and can be
measured with a high throughput technology \cite{Florian}. Protein
levels will also depend on the input signal. The proteins can be
recorded with 2D PAGE analysis or mass spectrometry. If one single
gene product is targeted, than a real-time luminescence recording
can be employed \cite{Yamazaki}. A periodic generator can be used
to investigate biological system for which the mRNA and protein
concentrations naturally oscillate in time. An example of such a
system is the circadian clock that drives a 24-hour rhythm in
living organisms from human to cyanobacteria. The core oscillator
is a molecular machinery based on an autoregulatory feedback loop
involving a set of key genes (Bmal, Per1,2,3, Cry1,2 , etc.)
\cite{Reppert}. Experimental procedures used to elucidate the
clock mechanism are based on measuring the circadian wheel-running
behavior of mice under normal light/dark (LD) cycles or in
constant darkness (dark/dark or DD) conditions. Experimental
evidence demonstrates that laws of quantitative nature govern the
molecular clock. For example, \cite{Gijbertus}, the internal clock
of cry 1 mutants have a free-running (i.e. DD conditions) period
of $22.51 \pm 0.06$ h which is significantly lower than the period
of a wild-type mice which is $23.77 \pm 0.07$ h. Quite opposite, a
cry 2 mutant have a significantly higher period of $24.63\pm 0.06$
h. In LD conditions, both mutants follow the 24 h period of the
entrained light cycles. A double cry1,2 mutant is arrhythmic  in
DD conditions and follow a 24 h rhythm in LD conditions. To
explain these experimental values we suggest using a light
switchable generator to drive the expression level of cry1, 2 and
measure the dynamics of transcription and the translation for the
rest of the key clock genes. Another application of the periodic
generator is to modulate a constitutively expressed gene by
superimposing an oscillatory profile on top of its flat level.
Then, the genes that show a modulation with a frequency equal to
the generator's frequency will be detected by a microarray
experiment. Why is this approach different from the one where a
step stimulus is used? Because the frequency of the generator is
{\it not} an internal parameter of the biological system. The
genes that interact with the driven gene will be modulated by the
input frequency. The rest of the genes will have different
expression profiles, dictated by the internal parameters of the
biological system. This point of view is supported by our
findings, \cite{Florian}, that the circadian clock (which is an
endogenous periodic signal generator) propagates its output to
only $8-10\%$ of the transcriptome in mice peripheral tissues
(liver or heart). In contrast to the oscillatory input, when a
step stimulus is applied, all the expression profiles are dictated
by the internal parameters of the biological system. Except for
the height of the step stimulus (the dose of the factor applied)
there is no external parameter implemented into the input signal.
As such, it is difficult to separate those genes that directly
respond to the input signal and to consequently avoid artifacts.
With the applications described in mind, we study the propagation
of an input signal through a stochastic genetic network.

\section{The response of a stochastic genetic network to an input
stimulus}

The effects of an oscillatory input were previously studied on
specific examples using models based on differential equations
\cite{Baxter},\cite{Jeff},\cite{Simpson}. The stochastic character
is embedded into these equations as an exterior additive term. In
contrast, we compute the generator's effects on the mean and
fluctuation of the gene products using a stochastic model
\cite{Mukund},\cite{Siggia},\cite{VanKampen}. In this way, the
generated stimulus and the noisy nature of the cell are entangled
in the stochastic genetic model. For a network of n genes the
state of a cell is described by the mRNA and protein molecule
numbers: $q=(r_1,...,r_n,p_1,...,p_n)$. We assume that, during any
small time interval $\Delta  t$, the probability for the
production of a molecule of the ith type is
$(\sum_{j=1}^{2n}A_{ij}q_j +G_i(t) ) \Delta t$, i.e. $q_i $ is
increased by 1 with the above probability. The function $G_i(t)$
represents the time varying input signal and modulates the mRNA
production only: $G=(g_1(t),\dots,g_n(t),0,\dots,0)^T$ (the
superscript $T$ is the transposition operation that transforms $G$
into a column vector for notational convenience in what follows).
The parameter $A_{ij}$ represents the influence of the jth type of
molecule on the production rate of a molecule of the ith type.
Similarly, there is a matrix of parameters $\Gamma_{ij}$ governing
the degradation rates of the molecules. For simplicity, we assume
that the input stimulus directly affects only the production
rates. The mean $\mu=\,\langle q \rangle \,$ and the covariance
matrix $\nu=\, \langle\langle q\rangle\rangle=\langle
 (q-\langle q\rangle)( q-\langle q\rangle) ^T\rangle\,$ of the state $q$ are driven
 by the generator $G$.

 The transfer of the signal from the generator through the genetic
network to the output measured data is encapsulated in a set of
transfer matrices. Specifically, let $H=A- \Gamma $ and denote the
Laplace transforms of $\mu $ and $G$ by ${\cal L} \mu$ and ${\cal
L} G$. Here and in what follows, $\mu $ and $G$ are represented as
column vectors. The connection between the mean and the generators
is given by formula (1) which is typical for a deterministic
linear system. However, the genetic system is stochastic and the
measure of the intrinsic noise is quantified by the covariance
matrix $\nu$. The effect of the stimulus generators is most
transparent if we split $\nu $ in a Poisson and a non-Poisson
component: $ \nu =diag(\mu) +X$. Here $diag(\mu)$ represents a
matrix with the components of the vector $\mu $ on its diagonal,
all the other terms being zero. For a Poisson process, $X=0$ and
thus the term $diag(\mu)$ is called the Poisson component of $\nu
$. The non-Poisson component $X= \nu -diag(\mu )$ can be expressed
in terms of the generators (Appendix and Supplementary Material):
%\end{multicols}
\begin{eqnarray}
 {\cal L}\mu &=& \frac{1}{(s-H)}{\cal L}G\;.\\ \label{Mean}
{\cal L}vec(X)&=&\frac{1}{s-1\otimes H-H\otimes 1}\left[\left(
{\it 1\otimes H}+{\it H\otimes 1} \right) L+2\,L{\it
   \Gamma}\right]\frac{1}{s-H} {\cal L}G\;.\label{GreenFunctionfFluctuation}
\end{eqnarray}
%\begin{multicols}{2}

\begin{figure}
\centering
\includegraphics[width=14cm]{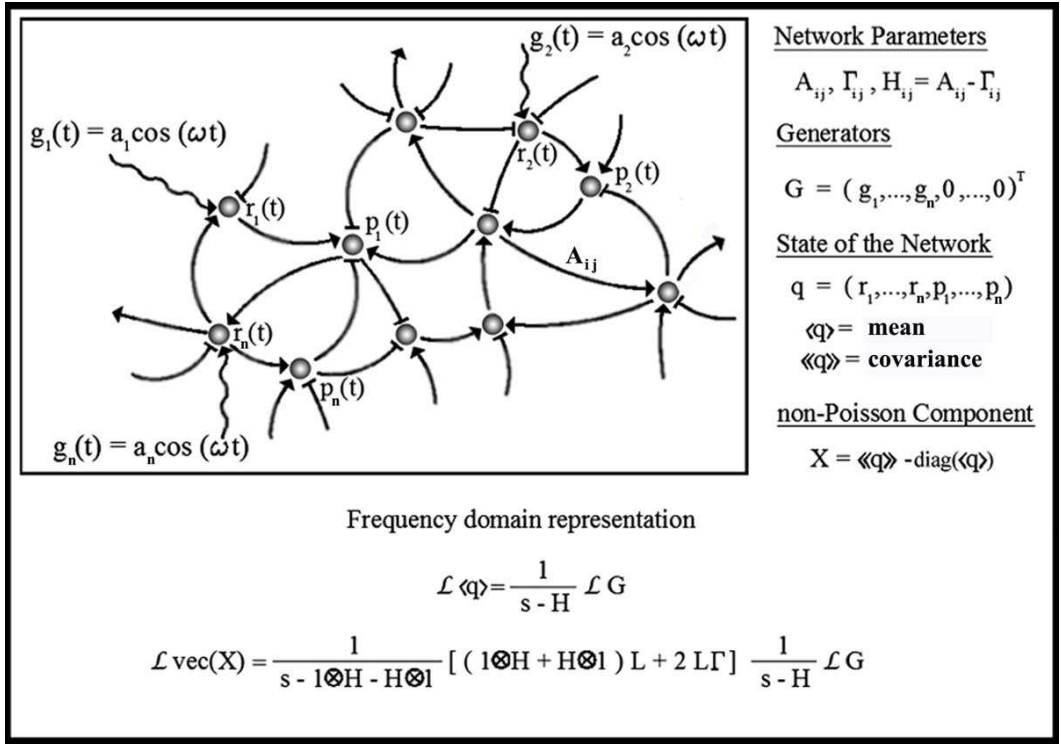}
\caption{Response of a stochastic genetic network to an
oscillatory input. The Laplace transform $\cal L$ change the
dynamic variable from time to frequency. In the $vec(X)$ all the
elements of the matrix X are arranged in a column vector.}
\end{figure}

The $vec(X)$ is a vector constructed from the matrix X by
stripping the columns of X one by one and stack them one on top of
each other in $vec(X)$. We emphasize here that the time variation
of the generators $G$ in (2) can take any form and is not bounded
to be periodic or a step stimulus.

 There are 3 matrices that transfer the information from the
generators to the non-Poisson component, ${\cal L}vec(X)=M_3 M_2
M_1 {\cal L}G.$ The first, $M_1=(s-H)^{-1}$, is the same as the
transfer matrix for the mean. The second, $M_2=(1\otimes
H+H\otimes 1)L+2L \Gamma $, breaks the symmetry between the
degradation and production parameters that are otherwise hidden in
the matrix $H=A-\Gamma $. The $ \otimes$ is the Kronecker product
of two matrices. The matrix $L$ (with elements $0$ and $1$) is the
lifting matrix  from dimension of the mean ($2n$ values) to the
dimension of $vec(X)$ ( $4n^2$ values). The third matrix is
$M_3=(s-1\otimes H - H\otimes 1)^{-1}.$ If $\lambda _i$ are the
eigenvalues of $H$ then all combinations $\lambda_i +\lambda_j$
are the eigenvalues of $1\otimes H+H\otimes 1$. Thus $M_3$
represents the analog of $M_1$ in the space of covariance
variables.

 For a
step stimulus, these eigenvalues are of primal importance: the
measured signal is a superposition of components with different
eigenvalues and has a complicated mathematical expression.
However, for a periodic stimulus, the frequency of the external
generator is the important parameter. This frequency is fixed by
the experimentalist not by the biological system. Only the phase
and the amplitude of the output signal depends on the system's
eigenvalues and the mathematical form is less cumbersome then for
the step stimulus. The input-output relations, (1) and (2), were
derived from the Master Equation written for the probability of
the states of the genetic network. Thus we must specify the
initial conditions for the probability of the states. These
conditions refer here to states for which one molecular component
vanishes ($ q_i=0$, for one $i$). The input-output relations, (1)
and (2), are independent of these boundary states if the $\Gamma $
matrix is diagonal. A diagonal $\Gamma $  matrix was used in
\cite{Mukund} and we will use it also in the example that follows.
Tools developed in the field of System Identification can be used
to create models for the networks under study, \cite{SystemId}.
The difference between the System Identification classical models
and a genetic network is that the later is a stochastic process by
nature, whereas the former are deterministic models with a
superimposed noise from external sources. However the formulas
that describe the relations between the mean and covariance of the
stochastic process and the input signals, (1) and (2), are of the
same general nature as those used in System Identification Theory,
\cite{SystemId}. In the next section we will use (1) and (2) to
analyze one of the most fundamental regulatory motifs in a genetic
network: an autoregulatory gene that acts upon itself through a
negative feedback, \cite{Young},\cite{Becskei},\cite{Rosenfeld}.
The fluctuation can drive this biological system out of its
equilibrium state,\cite{Isaacs}.

\section{Fluctuation resonance}

 Four parameters characterize the system:
the feedback strength $A_{12}=-h$ ; the translation rate
$A_{21}=k_p$ , and two degradation rates, $\Gamma_{11}=\gamma_r$,
$\Gamma_{22}=\gamma_p$. The gene regulation is under the control
of its own protein product and the protein activity is modulated
by a cofactor. The cofactor is driven by a periodic light
switchable generator $g(t)=k_0 + a cos(\omega t)$, Fig.1(a).
Before the generator is applied, the transcription rate is equal
with $k_0$ and the system is in a steady state. Through the
transfer matrices, (1) and (2), the light generator will impose a
periodic evolution of the mean and covariance matrix for mRNA and
protein product. We denote the mean mRNA by $\langle r(t) \rangle$
and the mean number of protein by $\langle p(t) \rangle$. We will
concentrate on the protein number in what follows. After the
transients are gone, $\langle
p(t)\rangle=P_{{0}}+P_{{1}}{e^{i\omega\,t}}+{ P}_{{1}}^{*}{e^{-i
\omega\,t}}$, that is the protein number will oscillate with an
amplitude $P_1$ on top of a baseline $P_0$; here $*$ represents
complex conjugation. The fluctuation of the protein number, $\,
\langle\langle p(t)\rangle\rangle \,$, differs from the mean
number by a quantity that we denoted by $X_{pp}(t)$:
$\,\langle\langle p(t)\rangle\rangle \,=\,\langle p(t) \rangle\,
+X_{pp}(t)$. For a pure Poisson process, $\,\langle\langle
p(t)\rangle\rangle \,=\,\langle p(t) \rangle$. Thus the term
$X_{pp}(t)$ represents the deviation from a Poisson process. If
there is some information about the genetic system that can be
uncovered by measuring not only the mean but also the covariance
matrix, then this information is hidden only in the non-Poisson
component $X_{pp}(t)$. The quantity $X_{pp}(t)$ is not interesting
only from a statistical point of view but also from a dynamical
one. The equation for the time evolution of $\, \langle\langle
p(t)\rangle\rangle \,$ takes its most simple form if it is written
for $X_{pp}(t)$. That is, the time dependence of the mean value
must be subtracted from the time evolution of $\, \langle\langle
p(t)\rangle\rangle \,$. Similar to the mean value, the non-Poisson
component of the fluctuation will oscillate in time, $X_{{pp}}(t)
=X_{{p,0}}+X_{{p,1}}{e^{i\omega\,t}}+{X}_{{p,1}
}^{*}{e^{-i\omega\,t}}$ with complex amplitude $X_{p,1}$. The
relative strength of the fluctuation versus the mean value can be
described using the Fano factor, \cite{Mukund}: $  {\,
\langle\langle p(t)\rangle\rangle \,}/{\,\langle p(t)
\rangle\,}=1+{X_{pp}(t)}/{\,\langle p(t) \rangle\,}.$ For
oscillatory inputs, the response of the network is best described
in frequency domain rather than in time.
 In frequency domain, as an analog of the Fano factor we consider
the ratio of the amplitude of $X_{pp}(t)$ versus the amplitude of
$\,\langle p(t) \rangle\,$.

%\end{multicols}
\begin{equation}\label{RatioResonance}
    \frac{\mid X_{p,1}\mid}{\mid P_1\mid}=\left(4\,{k_{{p}}}^{2}{\frac { \left( {\omega}^{2}+ \left( h-{\gamma_p} \right) ^{2} \right)
 \left( {\omega}^{2}+4\,{\gamma_{{r}}}^{2} \right) }{ \left(  \left( {
\omega}^{2}-4\,{\omega_{{0}}}^{2} \right)
^{2}+4\,{\omega}^{2}{\omega_ {{1}}}^{2} \right)  \left(
{\omega}^{2}+{\omega_{{1}}}^{2} \right) }}\right)^{1/2}\;.
\end{equation}
%\begin{multicols}{2}
Here $\omega_1=\gamma_r+\gamma_p$.
 The
complex amplitudes ${X}_{{p,1}}$ and $P_1$ depend on the input
frequency and therefore resonance phenomena can be detected in the
system. If the light switchable generator oscillates with double
the natural frequency $\omega_0^2=h k_p+\gamma_r \gamma_p$ , that
is , $\omega =2\omega_0$ we find a state of resonance for
fluctuation and not for the mean, Fig.3.

\begin{figure}
\centering
\includegraphics[width=14cm]{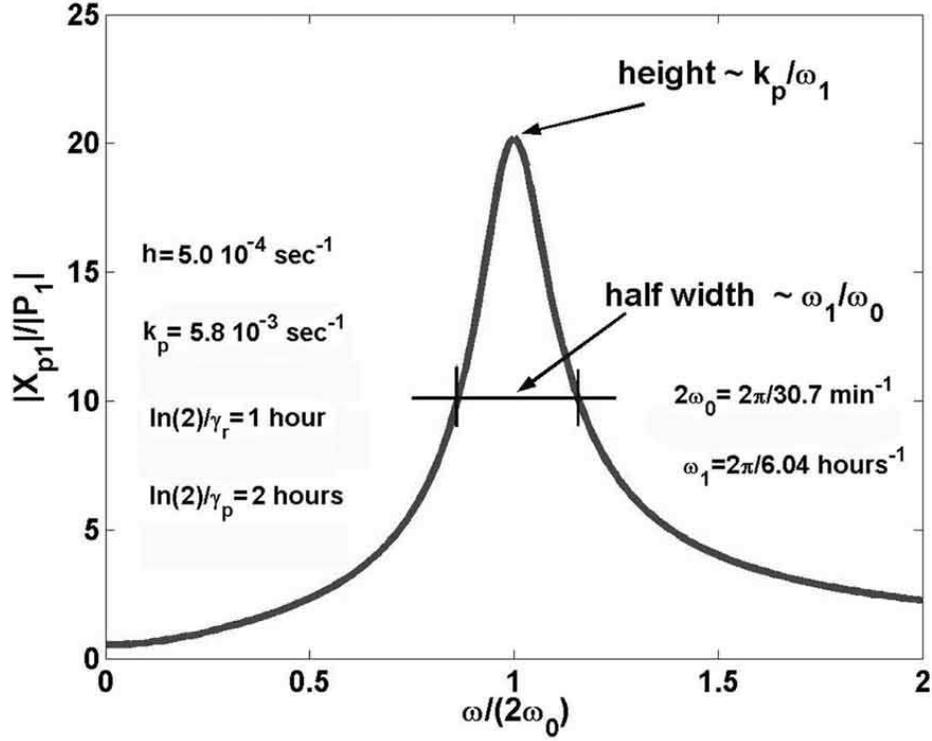}
\caption{Fluctuation resonance. The amplitude $X_{p,1}$ of the
non-Poisson component is much higher than the amplitude of the
mean protein number, $P_1$, at $\omega =2\omega_0.$}
\end{figure}

 For $\omega =2\omega_0$ the system will be in a pure fluctuation
resonance. In such a situation the molecular noise can drive the
cell out of its equilibrium state, which can have dramatic
consequence on the cell fate. Our model being linear cannot cover
the entire phenomena that accompanies a system whose state is
close to resonance. However, a linear model suggest the existence
of pure fluctuation resonance. At fluctuation resonance, the
deviation from a Poissonian process  is high. The oscillation
amplitude for protein fluctuation  is much greater then the
amplitude of the mean. Experimental results \cite{Kennell} show
that typical values for the ratio $k_p/\gamma _r$  are $40$  for
$lacZ$   and 5  for $lacA$. This suggests that there are natural
conditions for a strong height fluctuation resonance, Fig.3.
However, for a sharp fluctuation resonance (small half width), we
need  $h > \gamma _r$ or $\gamma _p$,  a condition that does not
appear in all genetic networks. It is with the help of the
experimental study that we will clarify why some biological
systems can sustain fluctuation resonance and others not. Beside
resonance, the frequency response provides other insights into the
structure of the autoregulatory system. The parameters of the
system can be read out from the measured data. The frequency
response of the mean values behave like the response of a
classical linear system to input signals. The new aspects are
those related to fluctuations. Like $X_{{pp}}(t)$ and
$X_{{rr}}(t)$, the correlation coefficient between the mRNA and
protein number will oscillate in time: $X_{{rp}}(t)
=X_{{rp,0}}+X_{{rp,1}}{e^{i\omega\,t}}+X_{{rp,1}
}^{*}{e^{-i\omega\,t}}$ with amplitude $X_{rp,1}$. Taking the
ratios of the amplitudes: ${ {   \left| X_{{rp,1}} \right|
^{2}}/{
 \left| X_{{r,1}} \right|
 ^{2}}}=({1}/{4h^2})\omega^2+{\gamma_r^2}/{h^2}\;,$
${{   \left| X_{{rp,1}} \right|   ^{2}}/{
 \left| X_{{p,1}} \right|
 ^{2}}}=({1}/{4k_p^2})\omega^2+{\gamma_p^2}/{k_p^2}\;,$
we observe that all four parameters of the system can be estimated
from the slopes and the intercepts of the above ratios as a
function of $\omega ^2$. Detail formulas for each amplitude are
given in the Supplementary Material.

\section{ The spectrum, the experimental noise and the importance of the
input stimulus }

We described the use of a periodic signal to decipher a genetic
network. Traditionally, a step stimulus is employed in biology for
pathway detection (i.e., adding a growth factor to the culture).
From the response to a step stimulus we can extract, in principle,
the parameters of the system.
 The natural question is then: why should we generate a
periodic stimulus when there is already a step stimulus in use?
Seeking an answer, we notice that the measured data in our studied
example can be expressed as a sum of exponentially decaying
functions, $ e^{-\lambda t}$, if a step stimulus was used
(Supplementary Material). For a periodic input, the response
contains only exponentials with imaginary argument, $ e^{i\omega
t}$. Mathematically, the main difference between exponentials with
real arguments, $e^{-\lambda t}$, and those with imaginary
arguments, $ e^{i \omega t}$, is that with the former we can not
form an orthogonal basis of functions whereas  such a basis can be
formed with the later. If we depart from our example, we can say
that in general, the response of the network to a step input will
be a sum of components which are not orthogonal on each other. The
time dependance of these nonorthogonal components can be more
complex than an exponential function; they can contain polynomials
in time or decaying oscillations, depending on the position in the
complex plane of eigenvalues of the transfer matrix $H$. Contrary,
the permanent response obtained from a periodic input is a sum of
Fourier components that form an orthogonal set. Orthogonal
components are much more easy to separate than nonorthogonal ones.
This mathematical difference explains the advantage of using
oscillatory inputs. However, an argument can be made that
increasing the number of replicates will be enough to recover the
step response form noise. In what follows we study how many
replicates we need to successfully fight the experimental noise.
We will show that we need fewer replicates if the genetic network
is probed with an oscillatory generator than with a step signal.
To keep the argument simple, we will study the difficulty of
separating  nonorthogonal components for a network for which the
response to a step stimulus is a sum of decaying exponentials. The
argument can be extended to other types of nonorthogonal
components, but this line of thought falls out of the scope of
this paper. The measured data being a superposition of exponential
terms can be written as:
\begin{equation}\label{Spec}
f \left( t \right) =\int _{x_1}^{x_2}\!S \left( x \right) K \left(
x\, t \right) {dx}\;,
\end{equation}

with $K(x t)=e^{-ix t}$ for the periodic response and $K(x
t)=e^{-x t}$ for the step stimulus.  The spectral function $S(x)$
depends on the network's parameters and on the type of the input
signal. For example, the spectrum of the autoregulatory system for
a periodic input is $S(x)=S_0\delta(x)+S_1\delta(x-i\omega)+{
S_1^{*}}\delta(x+i\omega)$, where $\delta(x)$ is the Dirac delta
function. The coefficients $S_0$, $S_1$ take specific values if
the spectrum refers to mean mRNA, proteins or their correlations.
For example, for the protein fluctuation:
\begin{eqnarray}
S_0=X_{{p,0}}&=&{\frac {{k_{{p}}}^{2}k_{{0}} \left( \gamma_{{p}}-h
\right) \gamma_{{r}}}{{\omega_{{0}}}^{4}\omega_{{1}}}}\;,
\\
S_1=X_{{p,1}}&=&{\frac {ia \left( -i\gamma_{{p}}+\omega+ih \right)
\left( \omega-2\,i \gamma_{{r}} \right) {k_{{p}}}^{2}}{ \left(
{\omega}^{2}-{\omega_{{0}} }^{2}-i\omega\,\omega_{{1}} \right)
\left( {\omega}^{2}-2\,i\omega\,
\omega_{{1}}-4\,{\omega_{{0}}}^{2} \right)  \left(
\omega-i\omega_{{1} } \right) }}\;.
\end{eqnarray}
Detailed description of the spectrum for an autoregulatory network
is given in the Section 5 of the Supplementary Material. For
oscillatory  inputs that are not pure cosine function and for more
complicated networks, the spectrum is more complex, but still is
connected with the measured data like in (\ref{Spec}). The
spectrum $S(x)$ carries information about the parameters of the
genetic network and it can be recovered from the data $f(t)$. The
network's parameter can be estimated from the spectrum once a
model of the network is chosen. Our goal is to show that the
spectrum obtained from an oscillatory input signal is much less
distorted by the experimental noise than the spectrum obtained
from a step input.
 Laboratory measurements
are samples of $f(t)$ at $N$ discrete time points. Given a finite
number $N$ of measured data points, $f_1,\cdots,f_N$, the spectrum
for the periodic case $S(x)$ can only be approximated as a
weighted sum of $N$ terms, (Supplementary Material): $ S(x)=\sum
_{k=1}^{N }(s_k+{\epsilon_k}/{\beta_k})\Theta_k(x).$ Each term,
$(s_k+{\epsilon_k}/{\beta_k})\Theta_k(x)$, contain a function
$\Theta_k(x)$ that do not depend on the measured data, and the
weights $s_k+ \epsilon_k / \beta_k$ that are computed from the
measured data $f_1,\cdots,f_N$. In the absence of experimental
noise, $\epsilon_k=0$, all $N$ coefficients $s_k$ can be computed
from the measured data. When experimental noise is present,
$\epsilon_k\neq 0$, what we compute from measured data is $s_k+
\epsilon_k / \beta_k$, and we cannot separate $s_k$ from it
because we do not know the actual value for $\epsilon_k$. The best
we can do is to use only those terms for which $s_k > \epsilon_k /
\beta_k$, so the effect of the distortion on $s_k$ is not large.
Unfortunately, the distortion increases as $\beta_k$ goes smaller,
which actually happens when $k$ increases. A term can be recovered
from noise if ${\beta_k}^{-1}< {s_k}/{\epsilon_k}$. Usually, this
relation is valid for $k=1\cdots J_p$, with $J_p$ being the last
term that can be recovered. A similar relation holds for the
exponential case, with $\alpha_k$ instead of $\beta_k$ and $J_e$
instead of $J_p.$ It is desirable that both cutoffs ($J_p,J_e$) be
as close as possible to the number of sampled points, $N$. The
striking difference between the two cases is that the cutoff $J_p$
is much larger then the cutoff $J_e$. This is a consequence of the
fact that the numbers $\alpha_k$ decrease exponentially to $0$,
\cite{Italy}, whereas $\beta_k$ stays close to 1 for many k before
eventually dropping close to zero, \cite{Slepian}. This huge
difference between $\alpha_k$ and $\beta_k$ has its origin in the
fact that the set of functions of time, $exp(-\lambda t)$, indexed
by $\lambda$, do not form an orthogonal set, whereas the functions
$exp(i\omega t)$, indexed by $\omega $, are orthogonal.
\begin{figure}[h]
\centering
\includegraphics[width=10cm]{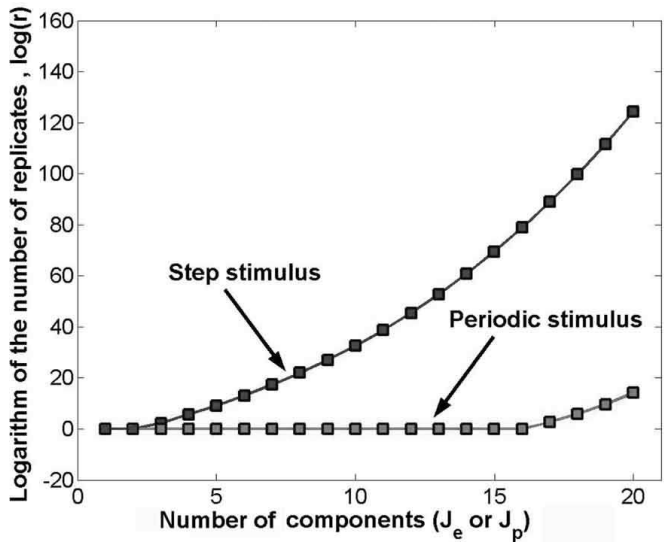}
\caption {How many replicates we need to recover a given spectral
component.}
\end{figure}
 In theory, however, we can still
hope that a step stimulus can deliver good estimates if the noise
$\epsilon_k$  is reduced  using $r$ replicates ($\epsilon_k$
$\rightarrow $ $\epsilon_k/\sqrt{r}$). This is not the case. Fig.4
represents the number of replicates needed to recover the
component $J_e$ or $J_p$ if the Signal to Noise Ratio is $10$
($SNR\equiv s_{J_e}/\epsilon_{J_e}=s_{J_p}/\epsilon_{J_p}=10$).
The number of replicates grows very fast in the exponential case
(for $SNR=10$ and $N=20$ we need 269 replicates for the 4th
spectral component), whereas in the periodic case, the number of
replicates stays low for many spectral components (only for the
17th component it raises to 14, with $SNR=10$ and $N=20$).

\section{Conclusions and Discussions}

We studied the response of a linear stochastic genetic network to
an input stimulus (signal). We provide a general formula that
relates the mean and covariance matrix of mRNAs and proteins to
the input generators. The particular type of periodic signals was
studied in detail for an autoregulatory system. We found that
fluctuation resonance can manifest in such systems. Besides
interesting physical phenomena that can be detected using a
periodic signal, the oscillatory input is useful for experimental
noise rejection. We compared two experimental designs: one that
uses a step stimulus as a perturbation and another one that uses a
periodic input. We concluded that the response of the genetic
network to a periodic stimulus is much easier to be detected from
noise than the response of the same network to a step stimulus.
This conclusion applies whenever the response of the network to an
oscillatory input is a sum of Fourier components. This can be the
case for many nonlinear networks.  However, the input-output
relations, (1) and (2), applies only to a linear stochastic model.
A linear model is a good approximation around a steady state of
the genetic network. A genetic network is a nonlinear system and
can have several steady states. If the signal generator does not
vary in time, the genetic network will be characterized by one of
these steady states. When the signal generator starts to oscillate
with an amplitude that doesn't drive the network far away from its
steady state, the linear model is a good approximation. For large
amplitudes, the nonlinear effects start to be important, and at
some values of the generator's amplitude, the network will jump
close to a different steady state. Such nonlinear behaviors can
not be described by a linear model. Also, the parameters that
describe the network are supposed to be constant in time. This
approximation is valid if the changes in the network parameters
are slow with respect to the changes produced by the oscillatory
input signals. The input frequency should be chosen so that the
system can be considered with constant coefficients for the
elapsed time of measurements. Also, the period of oscillations
must be less than the trend effects due to growth, apoptosis etc.
Beside biological effects that span large intervals of time,
experimental artifacts, like medium evaporation, can superimpose a
trend on the measured profile. The input period should be less
than the time characteristics of these trends. This will impose a
limit for the lower range of the input frequencies. The response
to oscillations depends also on the time characteristics of the
system under study. If the system has a high damping factor, the
high frequencies will be strongly attenuated and the output signal
is not measurable. With all these restrictions, the
experimentalist still has the freedom to work in a frequency band,
a freedom not present in the step stimulus.

 A different line of thought emerges when it comes to analyzing if
the oscillatory method can be scaled to large networks.
Experimentally, using high throughput measurements (microarray and
proteomic tools) a large set of gene products can simultaneously
be measured. The experimentalist is searching for a pathway that
is controlled by a gene. Using oscillatory signals to stimulate
the desired gene, the time variation of the downstream genes will
contain in its spectrum the input frequency and so these genes
will be detected. Moving the signal generator along the pathway,
more and more local patches of the network will be uncovered. The
global view on the network will consist of all these patches
connected together. The theoretical framework for connecting a set
of patches is unclear to us at present. Experimentally however, we
verified that a source of oscillations propagates into a large
genetic network,\cite{Florian}. Specifically, a microarray
experiment was conducted on mice entrained for two weeks on a 24
hours period of light-dark signals. The periodic input signal was
not implemented at the level of gene promoter; it was an exterior
periodic source of light that entrained the internal clock of the
cell. After entrainment, and in complete darkness, the output
signals (mRNA) were measured every 4 hours for 2 days using and
Affymetrix platform. From about 6000 expressed genes in heart,
about 500 showed a mRNA that oscillates with a 24 hours period.
Same results were reported in \cite{SanDiegoClock}. The next step
is to implement the generator at the gene promoter level, and
measure the spread of the input signal into the network.

Given the advantages of a periodic stimulus presented above, we
believe that the experimental implementation of a periodic
generator at promoter level will prove fruitful in the study of
genetic networks.

\appendix {{\bf Appendix}}

 The genetic network is described by a linear
stochastic network $\cite{Mukund},\cite{Siggia},\cite{VanKampen}$.
 The network is driven using signal generators placed inside the
 promoters of a subset of genes that are part of the network.
  For a gene we will denote by $(D,r,p)$ the number of DNA, mRNA and
protein molecules respectively per cell. We consider $r,p$
variables but $D$ constant and we normalize it to $D=1$. The state
of a cell that contains $n$ active genes is specified by: ${\tilde
q}=(D_1,D_2,...,D_n,r_1,r_2,...,r_n,p_1,p_2,...,p_n)$. The genetic
state is changing in time; for a short transition time, $dt$, only
one ${\tilde q}_i$ changes its value and this new value can be
either ${\tilde q}_i+1$ or ${\tilde q}_i-1$. We consider in this
paper a linear stochastic genetic network characterized by the
following transition probabilities: $T({\tilde q}_{};{\tilde
q}_{}+1_{i};t) = \sum_{j=1}^M {\tilde A}_{ij} {\tilde q}_j dt,$
 $T({\tilde q}_{};{\tilde q}{}-1_{i};t) =\sum_{j=1}^M{\tilde \Gamma}_{ij}{\tilde q}_j
 dt$. Here ${\tilde q}$ is the initial state and $1_i$ is a vector of
length M with all elements 0 except the one in the position $i$
which is 1. The time variation of the generators that drive the
genes' expressions are encapsulated in the matrix $\tilde A_{ij}$
which governs the production of different molecules. The matrices
$\tilde A_{ij}$ and $\tilde \Gamma_{ij} $ consist of  four
submatrices, corresponding to splitting the state $\tilde q$ in
two subgroups. One subgroup contains only the DNA states
$(D_1,\cdots,D_n)$ and the other subgroup contains the protein and
mRNA states $q=(r_1,r_2,...,r_n,p_1,p_2,...,p_n),$
\begin{equation}
\label{MatrixA} \tilde A=\left[ \begin {array}{cc}
0&0\\\noalign{\medskip}Gen&A\end {array} \right],\;\;\;
 \tilde \Gamma=\left[ \begin {array}{cc}
0&0\\\noalign{\medskip}0&\Gamma
\end {array} \right]\;.
\label{eqmu1}
\end{equation}
The generator submatrix $Gen$ has a special form. It is a $2n
\times n$ matrix and locates the position of the generators in the
genetic network:
$Gen_{ij}=g_i(t)\delta_{ij},\;i=1\dots2n,\;j=1\dots n.$
 Each gene promoter
is driven by one generator $g_i(t),i=1,\;\dots,n,$ which will
influence the mRNA production of gene $i$. The same mRNA
production can be influenced by the protein concentration, and
this feedback effect is described by the elements of the $2n\times
2n$ matrix $A$, (\ref{MatrixA}). The structure of the matrix $A$
is a consequence of the topology of the genetic network. The
equation for the probability $P({\tilde q},t)$ of the network to
be in the state ${\tilde q}$ at time $t$ is: $
 { {\partial P \left( \tilde {q},t \right)}/{\partial t}} =\sum
_{i=1}^{M} \left( {E_{{i}}}^{- }-1 \right) \sum _{k=1}^{M}\tilde
{A}_{{{\it ik}}}\tilde{q}_{{k}}P \left( \tilde {q},t \right) +
\sum _{i=1}^{M} \left( E_{{i}}^{+}-1 \right) \sum _{k=1}^{M}{\it
\tilde {\Gamma}}_{{ {\it ik}}}\tilde {q}_{{k}}P \left(\tilde {q},t
\right)\;, $
%\begin{multicols}{2}
where the shift operators $E_i^{\pm}$ are given by
$E_i^{\pm}P({\tilde {q}},t)=P({\tilde q}_1,...,{\tilde
q}_i\pm1,...,{\tilde q}_M)$.

We need the time evolution equations for mRNAs and proteins:
$\mu_i=< q_i>$ and $\nu_{ij}=<q_iq_j>-<q_i><q_j>$, $i,j=1,\dots
2n$. In matrix notation, for the column vector $\mu $ and for the
matrix $X$ with elements given by
$X_{ij}=\nu_{ij}-\delta_{ij}\mu_i$ we obtain:
%\end{multicols}
\begin{eqnarray}\label{EqMatrix1}
  {\frac {d}{dt}}\mu&=&H\mu+G \;,\\ \label{MatrixX1}
  {\frac {d}{dt}}X&=&HX+X{H}^{T}+ H {\it diag} \left( \mu \right) +{\it diag} \left(\mu \right)
 H ^{T}+2diag(\Gamma \mu)\;.\label{MatrixX2}
\end{eqnarray}
%\begin{multicols}{2}
 Here $H^T$ is the transpose matrix of $H=A-\Gamma$ and $diag(\mu)$
has nonzero elements only on the principal diagonal:
$diag(\mu)_{ij}=\delta_{ij}\mu_i$. Using the Laplace transform,
the solution to ($\ref{EqMatrix1}$) is (1). The second equation,
($\ref{MatrixX1}$), is a matrix equation. To solve this equation
we first transform it to an equation were the unknown is a column
vector. The transformation needed is $X\mapsto vec(X)$, where the
column vector $vec(X)$ contains the columns of the matrix $X$ one
on top of the next one, starting with the first column and ending
with the last column. The $vec$ mapping has the useful property
that $vec(HX)=(1\otimes H) vec(X)$, $vec(XH)=(H^T\otimes 1)
vec(X)$, were 1 is the unit matrix and $A\otimes B$ is the tensor
product of two matrices A and B. The column vector $vec({\it
diag}\left( \mu \right))$ can be expressed in terms of the column
vector $\mu$: $vec({\it diag}\left( \mu \right))=L\mu $, were $L$
is a lift matrix from a space of dimension of $\mu$ to the square
of this dimension: $L=(P_1,\dots,P_{2n})^T$,
$(P_k)_{ij}=\delta_{ik}\delta_{jk}$. The solution to
($\ref{MatrixX1}$) takes the form (2).

{\bf Acknowledgements}

We thank Kai-Florian Storch for valuable discussions about the
experimental design of the periodic generator and  Charles J.
Weitz for pointing to us the cry1,2 mutant experiments. This work
was supported in part by NIH grant 1R01HG02341 and NSF grant
DMS-0090166

\section{{\LARGE{\bf Supporting Material }}}

\subsection{Derivation of the time evolution equations for the mean
and fluctuation driven by signal generators}

  The genetic state ${\tilde
q}=(D_1,D_2,...,D_n,r_1,r_2,...,r_n,p_1,p_2,...,p_n)$ is changing
in time; let the state be ${\tilde q}_{initial}$ at time $t_1$ and
${\tilde q}_{final}$ at a later time $t_2$. The probability of
transition from the initial to the final state is, in the most
general case, a function of the initial state, the final state and
the times of transitions: $T({\tilde q}_{initial};{\tilde
q}_{final};t_1,t_2)$. Following a common hypothesis, the
transition probability is proportional with the transition time
$t_2-t_1$ if this is very short ($t_2=t_1+dt$ with $dt$ an
infinitesimal small quantity). The transition time being short
only one ${\tilde q}_i$ changes its value and this new value can
be either ${\tilde q}_i+1$ or ${\tilde q}_i-1$. We consider in
this paper a linear stochastic genetic network characterized by
the following transition probabilities: $T({\tilde q}_{};{\tilde
q}_{}+1_{i},;t,t+dt) = \sum_{j=1}^M {\tilde A}_{ij} {\tilde q}_j
dt,$
 $T({\tilde q}_{};{\tilde q}{}-1_{i};t,t+dt) =\sum_{j=1}^M{\tilde \Gamma}_{ij}{\tilde q}_j
 dt$. Here ${\tilde q}$ is the initial state and $1_i$ is a vector of
length M with all elements 0 except the one in the position $i$
which is 1. The equation for the probability of the network to be
in the state ${\tilde q}$ at time $t$, $P({\tilde q},t)$,  is
then, \cite{Mukund} \cite{VanKampen},:
\begin{equation}
\label{Probability}
 {\frac {\partial}{\partial t}}P \left( \tilde {q},t \right) =\sum
_{i=1}^{M} \left( {E_{{i}}}^{- }-1 \right) \sum _{k=1}^{M}\tilde
{A}_{{{\it ik}}}\tilde{q}_{{k}}P \left( \tilde {q},t \right) +
\sum _{i=1}^{M} \left( E_{{i}}^{+}-1 \right) \sum _{k=1}^{M}{\it
\tilde {\Gamma}}_{{ {\it ik}}}\tilde {q}_{{k}}P \left(\tilde {q},t
\right)\;, \label{Probability}
\end{equation}

where the shift operators $E_i^{\pm}$ are given by
\begin{equation}
E_i^{\pm}P({\tilde {q}},t)=P({\tilde q}_1,...,{\tilde
q}_i\pm1,...,{\tilde q}_M).
\end{equation}

We need to obtain the time evolution equations for $<{\tilde
q}_\alpha>$ and $<{\tilde q}_{\alpha} {\tilde q}_{\beta}>$,

\begin{equation}
<{\tilde q}_\alpha>\equiv\sum _{{\tilde q}_{{}}=0}^{\infty
}{\tilde q}_{{\alpha}}P \left( \tilde {q},t
 \right)\;,
\end{equation}
\begin{equation}
<{\tilde q}_\alpha {\tilde q}_\beta>\equiv\sum _{{\tilde
q}_{{}}=0}^{\infty }{\tilde q}_{{\alpha}}{\tilde q}_{{\beta}}P
\left( \tilde {q},t
 \right)\;.
\end{equation}

The easiest way to follow the computations is to use the
z-transform of a function, defined by:

\begin{equation}
 {\bf{Z}} \left(P \left( \tilde {q},t \right)
\right)= \sum _{{\tilde q}_{{1}}=0,...,{\tilde
q}_{{M}}=0}^{\infty}{{z_{{1}}}^{{\tilde q}_{{1}}}\ldots {z
_{{M}}}^{{\tilde q}_{{M}}}P \left( \tilde {q},t \right) }\;.
\end{equation}

The argument $z$ of the z-transform will be displayed using the
notation

\begin{equation}
F \left( z ,t\right) = {\bf{Z}}(P \left( \tilde {q},t \right)).
\end{equation}

 The  quantities of interest are
related with the z-transform through
%\begin{center}
%\fbox{
%\begin{minipage}{0.50\textwidth}
  \begin{eqnarray}
F_\alpha&=&<{\tilde q}_\alpha>\;,\\ \nonumber
 F_{\alpha\beta}&=&<{\tilde q}_\alpha
{\tilde q}_\beta>-\delta_{\alpha \beta}<{\tilde q}_\alpha>\;.
\end{eqnarray}
%\end{minipage}
%}
%\end{center}

where $\delta_{\alpha \beta}$ is  the Kronecker delta symbol which
is $0$ if $\alpha \neq \beta$ and $1$ if $\alpha =\beta$ and

\begin{eqnarray}
F_\alpha&=&{\frac {\partial }{\partial z_{{\alpha}}}}F \left( z,t
\right)\mid_{z_i=1,\;i=1...M}\;,\\
 F_{\alpha\beta}&=&{\frac {\partial
}{\partial z_{{\alpha}}\partial z_{{\beta}}}}F \left( z,t
\right)\mid_{z_i=1,\;i=1...M}\;.
\end{eqnarray}

The derivatives of  the z-transform are not directly related to
the covariance matrix:
\begin{equation}
\nu_{\alpha\beta}=<{\tilde q}_\alpha {\tilde q}_\beta>-<{\tilde
q}_\alpha><{\tilde q}_\beta>\;.
\end{equation}
However, the covariance matrix can be easily expressed in terms of
the z-transform variables:
\begin{equation}
\nu_{\alpha\beta}=F_{\alpha\beta}-F_\alpha
F_\beta+\delta_{\alpha\beta}F_\alpha\;.
\end{equation}

The equation for $F$ can be obtained by taking the z-transform of
the master equation (\ref{Probability}) using the following rules:
\begin{eqnarray}\label{ZtransformRules}
{\bf Z}\left (E_i^{+}P({\tilde {q}},t)\right)&=& z_i^{-1}{\bf Z}\left (P({\tilde q},t) \right )-z_i^{-1}{\bf Z}\left( P({\tilde q},t)\mid_{{\tilde q}_i=0}\right ) \;,\\
{\bf Z}\left (E_i^{-}P({\tilde {q}},t)\right)&=& z_i{\bf Z}\left (P({\tilde q},t) \right ) \;,\\
{\bf Z}\left ({\tilde q}_iP({\tilde {q}},t) \right
)&=&z_i\partial_{z_i}{\bf Z}\left (P({\tilde q},t) \right ).\\
\end{eqnarray}

If the degradation matrix $\Gamma$ is diagonal \cite{Mukund}, then
the probability $P({\tilde q},t)\mid_{{\tilde q}_i=0}$ of the
state with a missing molecular specie will not be part of the the
equation for the z-transform. Indeed, the boundary term in the
z-transform of $E_i^{+}\Gamma_{ii}{\tilde q}_iP({\tilde q},t)$
will vanish for ${\tilde q}_i=0$. For a non-diagonal $\Gamma $
matrix, we obtain the same equation if we work with natural
boundary conditions, that is  $P({\tilde {q}},t)=0$ if ${\tilde
q}_i=0$ for one $i$ from the set $1...M$. The majority of the
genetic networks will not obey the natural boundary conditions.
However, the final results are the same for (i) a non-diagonal
$\Gamma $ matrix with natural boundary conditions and (ii) a
diagonal $\Gamma $ matrix with no restriction imposed on the
boundary. For the sake of the symmetry of the computations, we
will derive the results for a general $\Gamma $ matrix and natural
boundary conditions, and use a diagonal $\Gamma $ matrix when we
will study the behavior of a genetic network.

The equation for the z-transform now reads:
\begin{equation}\label{E1}
   {\frac {\partial }{\partial t}}F \left( z,t \right) =\sum _{i=1}^{M}
 \left( z_{{i}}-1 \right) \sum _{k=1}^{M}\tilde {A}_{{{\it ik}}}z_{{k}}{\frac {
\partial }{\partial z_{{k}}}}F \left( z,t \right) +\sum _{i=1}^{M}
 \left( {z_{{i}}}^{-1}-1 \right) \sum _{k=1}^{M}{\it \tilde {\Gamma}}_{{{\it ik}
}}z_{{k}}{\frac {\partial }{\partial z_{{k}}}}F \left( z,t
\right)\;.
\end{equation}

Take the derivative of these equation with respect to $z_\alpha $

\begin{eqnarray}
\nonumber
    {\frac {\partial^2 }{\partial t\partial z_{{\alpha}}}}F \left( z,t
    \right)=\sum _{{\it i,k}=1}^{M}\tilde {A}_{{{\it ik}}} \left( \delta_{{i\alpha}}z_{{k}}
{\frac {\partial }{\partial z_{{k}}}}F \left( z,t \right) + \left(
z_{ {i}}-1 \right) \delta_{{k\alpha}}{\frac {\partial }{\partial
z_{{k}}} }F \left( z,t \right) + \left( z_{{i}}-1 \right)
z_{{k}}{\frac {
\partial^2 }{\partial z_{{k}}\partial z_{{\alpha}}}}F \left( z,t
\right)
 \right)+\\
 \sum _{{\it i,k}=1}^{M}{\it \tilde {\Gamma}}_{{{\it ik}}} \left( -{z_{{i}}}^{-2}
\delta_{{i\alpha}}z_{{k}}{\frac {\partial }{\partial z_{{k}}}}F
 \left( z,t \right) + \left( {z_{{i}}}^{-1}-1 \right) \delta_{{k\alpha
}}{\frac {\partial }{\partial z_{{k}}}}F \left( z,t \right) +
\left( { z_{{i}}}^{-1}-1 \right) z_{{k}}{\frac {\partial^2
}{\partial z_{{k}}\partial z_{ {\alpha}}}}F \left( z,t \right)
\right)\nonumber\;.
\end{eqnarray}

Introducing $z_i=1,i=1...M$ we obtain the equation for the time
evolution of the mean values:

%\begin{center}
%\fbox{
%\begin{minipage}{0.42\textwidth}
  \begin{equation}
{\frac {d}{dt}}F_{{\alpha}}=\sum _{k=1}^{M}(\tilde {A}_{{\alpha
k}}-{\it \tilde {\Gamma}}_{{\alpha k}})F_{{k}}\;.
\end{equation}
%\end{minipage}
%}
%\end{center}

For the second moments we continue to take derivatives of
(\ref{E1}):

\begin{eqnarray}
\nonumber
  {\frac {\partial^3 }{\partial t \partial z_{{\alpha}}\partial z_{{\beta}}}}F \left( z,t
 \right)=
  \sum _{i,k=1}^{M}\tilde {A}_{{{\it ik}}} ( \delta_{{i
  \alpha}}\delta_{{k
\beta}}{\it \partial }_{{k}}F+\delta_{{i \alpha}}z_{{k}}{\it
\partial }_{{k \beta}}F+ \delta_{{i \beta}}\delta_{{k \alpha}}{\it
\partial }_{{k}}F+ \left( z_{{i}}-1
 \right) \delta_{{k \alpha}}{\it \partial }_{{k \beta}} F+\\
 \nonumber
 \delta_{{i \beta}}z_{{k}}\partial_{{k \alpha}}F+ \left( z_{{i}}-1 \right)
\delta_{{k \beta}}\partial_{{k \alpha}}F+ \left( z_{{i}}-1 \right)
z_{{k}}\partial_{ {k \alpha \beta}}F)+
  \\ \nonumber
  \sum _{{\it i,k}=1}^{M}{\it \tilde {\Gamma}}_{{{\it ik}}} ( 2\,{z_{{i}}}^{-3}
\delta_{{i\beta}}\delta_{{i\alpha}}z_{{k}}\partial_{{k}}F-{z_{{i}}}^{-2}
\delta_{{k\beta}}\delta_{{i\alpha}}\partial_{{k}}F-{z_{{i}}}^{-2}\delta_{{i
\alpha}}z_{{k}}\partial_{{k\beta}}F-  \\ \nonumber
  {z_{{i}}}^{-2}\delta_{{i\beta}}\delta_{{k\alpha}}\partial_{{k}}F+ \left( {z
_{{i}}}^{-1}-1 \right)
\delta_{{k\alpha}}\partial_{{k\beta}}F-{z_{{i}}}^{-2}
\delta_{{i\beta}}z_{{k}}\partial_{{k\alpha}}F+
 \\ \nonumber
  \left( z_{{i}}^{-1}-1 \right) \delta_{{k\beta}}\partial_{{k\alpha}}F+ \left( {z_{{i}}}^{-1}-1 \right)
  z_{{k}}\partial_{{k\alpha \beta}}F)\,,
\end{eqnarray}

\begin{eqnarray}
\nonumber
  {\frac {\partial^3 }{\partial t \partial z_{{\alpha}}\partial z_{{\beta}}}}F \left( z,t
 \right)\mid_{z_i=1,i=1..M}=\tilde {A}_{{\alpha \beta}}\partial_{{\beta}}F+\sum _{k=1}^{M}\tilde {A}_{{\alpha k}}\partial_{{k
\beta}}F+\tilde {A}_{{\beta \alpha}}\partial_{{\alpha}}F+\sum
_{k=1}^{M}\tilde {A}_{{\beta k}}\partial _{{k \alpha}}F+
  \\ \nonumber
\sum _{k=1}^{M}2\,{\it \tilde {\Gamma}}_{{\beta
k}}\partial_{{k}}F\delta_{{\alpha \beta }}-{\it \tilde
{\Gamma}}_{{\alpha \beta}}\partial_{{\beta}}F-\sum _{k=1}^{M}{\it
\tilde {\Gamma}} _{{\alpha k}}\partial_{{k \beta}}F-{\it \tilde
{\Gamma}}_{{\beta \alpha}}\partial_{{\alpha}}F- \sum
_{k=1}^{M}{\it \tilde {\Gamma}}_{{\beta k}}\partial_{{k
\alpha}}F\,,
\end{eqnarray}

$\;\;$

 $\;\;$

$\;\;$

\fbox{
\begin{minipage}{0.92\textwidth}
  \begin{eqnarray}\label{Fab}
{\frac {d}{dt}}F_{{\alpha \beta}}=\sum _{k=1}^{M} \left( \tilde
{A}_{{\alpha k} }-{\it \tilde {\Gamma}}_{{\alpha k}} \right) F_{{k
\beta}}+\sum _{k=1}^{M}
 \left( \tilde {A}_{{\beta k}}-{\it \tilde {\Gamma}}_{{\beta k}} \right) F_{{k \alpha}}+\tilde {A}
_{{\alpha \beta}}F_{{\beta}}+\tilde {A}_{{\beta
\alpha}}F_{{\alpha}}
 \\ \nonumber
  -{\it \tilde {\Gamma}}_{{\alpha \beta}}F_{{\beta}}-{\it \tilde {\Gamma}}_{{\beta \alpha}}F
_{{\alpha}}+2\,\delta_{{\alpha \beta}}\sum _{k=1}^{M}{\it \tilde
{\Gamma}}_{{ \beta k}}F_{{k}}\,.
\end{eqnarray}
\end{minipage}
}

$\;\;$

 $\;\;$

 This is the equation that we need. Later we will use it to
reveal the action of the generators, that are hidden now in the
coefficients $\tilde {A}_{ik}.$ Before we deal with the
generators, we will derive a general formula for the covariance
matrix $\nu_{\alpha \beta}$ to see how different it is from the
one above.

\begin{equation}\label{Miu}
\nu_{\alpha \beta}\equiv <{\tilde q}_\alpha {\tilde q}_\beta
>-<{\tilde q}_\alpha
><{\tilde q}_\beta
>=F_{\alpha \beta}-F_{\alpha}F_{\beta}+\delta_{\alpha
\beta}F_{\alpha}\;,
\end{equation}

\begin{equation}
   {\frac {d}{dt}}\nu_{{\alpha \beta}}={\frac {d}{dt}}F_{{\alpha \beta}}-
 \left( {\frac {d}{dt}}F_{{\alpha}} \right) F_{{\beta}}-F_{{\alpha}}{
\frac {d}{dt}}F_{{\beta}}+\delta_{{\alpha \beta}}{\frac
{d}{dt}}F_{{ \alpha}}\;.
\end{equation}

Now we insert the derivatives for $F_\alpha$ and $F_{\alpha
\beta}$

\begin{eqnarray}
{\frac {d}{dt}}\nu_{{\alpha \beta}}=\sum _{k=1}^{M} \left( \tilde
{A}_{{\alpha k} }-{\it \tilde {\Gamma}}_{{\alpha k}} \right) F_{{k
\beta}}+\sum _{k=1}^{M}
 \left( \tilde {A}_{{\beta k}}-{\it \tilde {\Gamma}}_{{\beta k}} \right) F_{{k \alpha}}+\tilde {A}
_{{\alpha \beta}}F_{{\beta}}+\tilde {A}_{{\beta
\alpha}}F_{{\alpha}}
 \\ \nonumber
  -{\it \tilde {\Gamma}}_{{\alpha \beta}}F_{{\beta}}-{\it \tilde {\Gamma}}_{{\beta \alpha}}F
_{{\alpha}}+2\,\delta_{{\alpha \beta}}\sum _{k=1}^{M}{\it \tilde
{\Gamma}}_{{ \beta k}}F_{{k}}+\\ \nonumber \sum _{k=1}^{M}(-\tilde
{A}_{{\alpha k}}F_{{k}}F_{{\beta}}+{\it \tilde {\Gamma}}_{{\alpha
k}}F_{{k}}F_{{\beta}}-\tilde {A}_{{\beta
k}}F_{{k}}F_{{\alpha}}+{\it \tilde {\Gamma}}_{{ \beta
k}}F_{{k}}F_{{\alpha}})+\\ \nonumber \delta_{{\alpha \beta}}\sum
_{k=1}^{M} \left( \tilde {A}_{{\alpha k}}-{\it \tilde
{\Gamma}}_{{\alpha k}} \right) F_{{k}}\nonumber \;.
\end{eqnarray}

We want to get rid of the variables $F_{\alpha\beta}$ and write
everything in terms of $\nu_{\alpha \beta}$ and $<q_\alpha >$.
First we regroup the terms and then add and subtract the term

\begin{equation}
   \sum _{k=1}^{M} \left( \tilde {A}_{{\alpha k}}-{\it \tilde {\Gamma}}_{{\alpha k}}
 \right) \delta_{{k \beta}}F_{{\beta}}
\end{equation}

to obtain
\begin{eqnarray}
  {\frac {d}{dt}}\nu_{{\alpha \beta}}=\sum _{k=1}^{M} \left( \tilde {A}_{{\alpha k}}-{\it \tilde {\Gamma}}_{{\alpha k}}
 \right)  \left( F_{{k \beta}}-F_{{k}}F_{{\beta}}+\delta_{{k \beta}}F_
{{\beta}} \right) -\sum _{k=1}^{M} \left( \tilde {A}_{{\alpha
k}}-{\it \tilde {\Gamma}}_{ {\alpha k}} \right) \delta_{{k
\beta}}F_{{\beta}}+
 \\ \nonumber
  +\sum _{k=1}^{M} \left( \tilde {A}_{{\beta k}}-{\it \tilde {\Gamma}}_{{\beta, k}} \right)
 \left( F_{{k \alpha}}-F_{{k}}F_{{\alpha}}+\delta_{{k \alpha}}F_{{
\alpha}} \right) -\sum _{k=1}^{M} \left( \tilde {A}_{{\beta
k}}-{\it
\tilde {\Gamma}}_{{ \beta k}} \right) \delta_{{k \alpha}}F_{{\alpha}}+\\
\nonumber +\tilde {A}_{{\alpha \beta}}F_{{\beta}}+\tilde
{A}_{{\beta \alpha}}F_{{\alpha}}-{\it \tilde {\Gamma}}_{{\alpha
\beta}}F_{{\beta}}-{\it \tilde {\Gamma}}_{{\beta \alpha}}F_{{
\alpha}}+2\,\delta_{{\alpha \beta}}\sum _{k=1}^{M}{\it \tilde
{\Gamma}}_{{\beta  k}}F_{{k}}+\\ \nonumber +\delta_{{\alpha
\beta}}\sum _{k=1}^{M} \left( \tilde {A}_{{\alpha k}}-{\it \tilde
{\Gamma}}_{{\alpha k}} \right) F_{{k}}
\end{eqnarray}

\begin{eqnarray}
  {\frac {d}{dt}}\nu_{{\alpha \beta}}=\sum _{k=1}^{M} \left( \tilde {A}_{{\alpha k}}-{\it \tilde {\Gamma}}_{{\alpha k}}
 \right) \nu_{{k \beta}}+\sum _{k=1}^{M} \left( \tilde {A}_{{\beta k}}-{\it
\tilde {\Gamma}}_{{\beta k}} \right) \nu_{{k \alpha}}-
 \\ \nonumber
 -\tilde {A}_{{\alpha \beta}}F_{{\beta}}+{\it \tilde {\Gamma}}_{{\alpha \beta}}F_{{\beta}}
-\tilde {A}_{{\beta \alpha}}F_{{\alpha}}+{\it \tilde
{\Gamma}}_{{\beta \alpha}}F_{{\alpha }}+
\\ \nonumber
+\tilde {A}_{{\alpha \beta}}F_{{\beta}}+\tilde {A}_{{\beta
\alpha}}F_{{\alpha}}-{\it \tilde {\Gamma}}_{{\alpha
\beta}}F_{{\beta}}-{\it \tilde {\Gamma}}_{{\beta \alpha}}F_{{
\alpha}}+
\\ \nonumber
+\delta_{{\alpha \beta}}\sum _{k=1}^{M}{\it \tilde
{\Gamma}}_{{\alpha k}}F_{{k}}+ \delta_{{\alpha \beta}}\sum
_{k=1}^{M}\tilde {A}_{{\alpha k}}F_{{k}}\nonumber
\end{eqnarray}

\fbox{
\begin{minipage}{0.98\textwidth}
  \begin{eqnarray}\label{EqMu}
{\frac {d}{dt}}\nu_{{\alpha \beta}}=\sum _{k=1}^{M} \left( \tilde
{A}_{{\alpha k}}-{\it \tilde {\Gamma}}_{{\alpha k}}
 \right) \nu_{{k \beta}}+\sum _{k=1}^{M} \left( \tilde {A}_{{\beta k}}-{\it
\tilde {\Gamma}}_{{\beta k}} \right) \nu_{{k
\alpha}}+\delta_{{\alpha \beta}}\sum _{k=1}^{M} \left( \tilde
{A}_{{\alpha k}}+{\it \tilde {\Gamma}}_{{\alpha k}} \right)
<{\tilde q}_{{k}}>\;\;\;\;\;\;
\end{eqnarray}
\end{minipage}
}

\subsection{  The Generators }

The generators constitute a submatrix of the matrix $\tilde {A}:$

\begin{equation}
\tilde {A}=\left(\begin{array}{cc}
  0_{\alpha \beta} & 0_{\alpha b} \\
  {\cal G}_{a \beta}  & A_{a b}\;.\\
\end{array}\right)
\end{equation}

Here $0_{\alpha \beta}$ and $0_{\alpha b}$ are matrices  with all
elements zeros, where $\alpha ,\beta =1\dots n$, and  $a,b=n+1
\dots 3n$. The matrix ${\cal G}_{a \beta}$ contains the generators
and thus is a matrix with time dependant elements. The matrix
$A_{a b}$ has constant elements which depend on the genetic
network. From now on we make a distinction between Greek indices
and Latin indices, so that we can rewrite the general time
dependance equations in terms of generators. The Greek indices run
along the DNA variables, whereas the Latin indices run through the
mRNAs and proteins variables.

Lets specialize the equation (\ref{Fab}) for Latin indices. We
will split the summations using a generic Greek letter $\gamma $
and a generic Latin letter $g$. We consider the number of DNA be
constant in time and normalized to the value 1. As a consequence

\begin{equation}
F_{\gamma b}\equiv <{\tilde q}_\gamma {\tilde q}_b>-\delta_{\gamma
b}<{\tilde q}_b>=<{\tilde q}_\gamma {\tilde q}_b>=<1 {\tilde
q}_b>=<{\tilde q}_b>=F_{b}
\end{equation}

In terms of Greek and Latin indices, the matrix $\tilde {\Gamma
}$, looks like

\begin{equation}
\left(\begin{array}{cc}
  0_{\alpha \beta} & 0_{\alpha b} \\
  0_{a \beta}  & \Gamma_{a b}\\
\end{array}\right)\;,
\end{equation}

so
\begin{eqnarray}
 \nonumber
  {\frac {d}{dt}}F_{{ab}}= \sum _{\gamma}^{} \left( A_{{a\gamma}}-{\it \Gamma}_{{a\gamma}}
 \right) F_{{\gamma b}}+\sum _{g}^{} \left( A_{{ag}}-{\it \Gamma}_{
{ag}} \right) F_{{gb}}+\sum _{\gamma}^{} \left( A_{{b\gamma}}-{\it
\Gamma}_{{b\gamma}}
 \right) F_{{\gamma a}}+\sum _{g}^{} \left( A_{{bg}}-{\it \Gamma}_{
{b g}} \right) F_{{ga}}+
\\ \nonumber
+A_{{{\it ab}}}F_{{b}}+A_{{{\it ba}}}F_{{a}}-{\it \Gamma}_{{{\it a
b}}}F_ {{b}}-{\it \Gamma}_{{{\it ba}}}F_{{a}}+2\,\delta_{{{\it a
b}}} \left( \sum _{\gamma}^{}{\it
\Gamma}_{{b\gamma}}F_{{\gamma}}+\sum _{g}^{ }{\it
\Gamma}_{{bg}}F_{{g}} \right)\,,\nonumber
\end{eqnarray}

\begin{eqnarray}
 \nonumber
  {\frac {d}{dt}}F_{{ab}}=(\sum _{\gamma}^{}{\cal G}_{{a\gamma}})<q_{{b}}>+(\sum _{\gamma}^{}{\cal G}_{{b
\gamma}})<q_{{a}}>+\sum _{g}^{} \left( A_{{ag}}-{\it
\Gamma}_{{ag}}
 \right) F_{{gb}}+\sum _{g}^{} \left( A_{{bg}}-{\it \Gamma}_{{bg}
} \right) F_{{ga}}+\\ \nonumber +A_{{{\it ab}}}F_{{b}}+A_{{{\it
ba}}}F_{{a}}-{\it \Gamma}_{{{\it ab}}}F_ {{b}}-{\it \Gamma}_{{{\it
ba}}}F_{{a}}+2\,\delta_{{{\it ab}}}\sum _{g }^{}{\it
\Gamma}_{{bg}}F_{{g}}\,. \nonumber
\end{eqnarray}

We have to eliminate the sum of the generators. We use for this
the equation for the mean, taking care that for DNA variables,
$F_\gamma =<{\tilde q}_\gamma
>=1$

\begin{eqnarray}
\label{UsefulFor The Mean}
  {\frac {d}{dt}}F_{{a}}=\sum _{\gamma}^{} \left( A_{{a\gamma}}-{
\it \Gamma}_{{a\gamma}} \right) F_{{\gamma}}+\sum _{g}^{} \left(
A_ {{ag}}F_{{g}}-{\it \Gamma}_{{ag}} \right) F_{{g}}\\
\nonumber =\sum _{\gamma}^{}{\cal G}_{{a\gamma}}+\sum _{g}^{}
\left( A_{{ag}}-{ \it \Gamma}_{{ag}} \right) F_{{g}}\;.
\end{eqnarray}

We obtain then:

\begin{eqnarray}
 \nonumber
  {\frac {d}{dt}}F_{{ab}}= \left( {\frac {d}{dt}}F_{{a}}-\sum _{g}^{} \left( A_{{ag}}-{\it \Gamma}_{{ag}} \right) F_{{g}} \right) F_{{b}}+ \left( {\frac {d}{dt}}
F_{{b}}-\sum _{g}^{} \left( A_{{bg}}-{\it \Gamma}_{{bg}} \right) F
_{{g}} \right) F_{{a}}+\\ \nonumber +\sum _{g}^{} \left(
A_{{ag}}-{\it \Gamma}_{{ag}} \right) F_{{gb}} +\sum _{g}^{} \left(
A_{{bg}}-{\it \Gamma}_{{bg}} \right) F_{{ga}
}+A_{{ab}}F_{{b}}+A_{{ba}}F_{{a}} -\\\nonumber -{\it
\Gamma}_{{{\it ab}}}F_ {{b}}-{\it \Gamma}_{{{\it
ba}}}F_{{a}}+2\,\delta_{{{\it ab}}}\sum _{g }^{}{\it
\Gamma}_{{bg}}F_{{g}}\;,
\end{eqnarray}

\begin{eqnarray}
\nonumber
  {\frac {d}{dt}}F_{{ab}}=\left( {\frac {d}{dt}}F_{{a}} \right) F_{{b}}+ \left( {\frac {d}{dt}}F_{{b}} \right) F_{{a}}+\sum _{g}^{} \left( A_{{ag}}-{\it \Gamma}_{
{ag}} \right)  \left( F_{{gb}}-F_{{g}}F_{{b}} \right) +\sum _{g}
\left( A_{{bg}}-{\it \Gamma}_{{bg}} \right)  \left( F_{{ga}}-F_{{
g}}F_{{a}} \right)+
 \\ \nonumber
 +A_{{ab}}F_{{b}}+A_{{ba}}F_{{a}}-{\it \Gamma}_{{{\it ab}}}F_ {{b}}-{\it \Gamma}_{{{\it
ba}}}F_{{a}}+2\,\delta_{{{\it ab}}}\sum _{g }^{}{\it
\Gamma}_{{bg}}F_{{g}}\;.
\end{eqnarray}

From the formula above, we see that a new variable appeared in a
natural way:
\begin{equation}\label{X}
    X_{ab}=F_{ab}-F_aF_b\;.
\end{equation}

The time evolution of this new variable is given by the equation:

%\begin{minipage}{0.98\textwidth}
  \begin{eqnarray}\label{EqX}
{\frac {d}{dt}}X_{{ab}}=\sum _{g}^{}(H_{{ag}}X_{{gb}}+H_{{bg}}X_
{{ga}})+A_{{ab}}<{\tilde q}_{{b}}>+A_{{ba}}<{\tilde q}_{{a}}>-\\
\nonumber -{\it \Gamma}_{{{\it ab}}}<{\tilde q}_ {{b}}>-{\it
\Gamma}_{{{\it ba}}}<{\tilde q}_{{a}}>+2\,\delta_{{{\it ab}}}\sum
_{g }^{}{\it \Gamma}_{{bg}}<{\tilde q}_{{g}}>\;,
\end{eqnarray}
%\end{minipage}

with
\begin{equation}\label{H}
    H_{ab}=A_{ab}-\Gamma_{ab}\;,
\end{equation}

or

%\fbox{
%\begin{minipage}{0.98\textwidth}
  \begin{eqnarray}\label{EqX}
{\frac {d}{dt}}X_{{ab}}=\sum _{g}^{}(H_{{ag}}X_{{gb}}+H_{{bg}}X_
{{ga}})+H_{{ab}}<{\tilde q}_{{b}}>+H_{{ba}}<{\tilde
q}_{{a}}>+2\,\delta_{{{\it ab}}}\sum
_{g }^{}{\it \Gamma}_{{bg}}<{\tilde q}_{{g}}>\;.\\
\nonumber
\end{eqnarray}
%\end{minipage}
%}

 In what follows we will use a diagonal $\Gamma$ matrix.
For this case the equation simplifies to

%\fbox{
%\begin{minipage}{0.98\textwidth}
  \begin{eqnarray}\label{EqX}
{\frac {d}{dt}}X_{{ab}}=\sum _{g}^{}(H_{{ag}}X_{{gb}}+H_{{bg}}X_
{{ga}})+A_{{ab}}<{\tilde q}_{{b}}>+A_{{ba}}<{\tilde q}_{{a}}>\;.\\
\nonumber
\end{eqnarray}
%\end{minipage}
%}

The meaning of the matrix $X$ can be found if we write it in terms
of the covariance matrix $\nu_{ab}$.

\begin{equation}\label{XMeaning}
    X_{ab}=F_{ab}-F_aF_b=\nu_{ab}-\delta_{ab}F_a=\nu_{ab}-\delta_{ab}<{\tilde
    q}_a>\;.
\end{equation}

Thus $X$ measure the deviation of the stochastic process from a
Poissonian process,

\begin{equation}\label{MuX}
    \nu_{ab}=\delta_{ab}<{\tilde q}_a>+X_{ab}\;.
\end{equation}

Now it is easy to write everything in terms of the reduced state
$q=(r_1,r_2,...,r_n,p_1,p_2,...,p_n).$ To do this we observe that
$q_k={\tilde q}_{k+n}$, $k=1\dots 2n$. In other words, we subtract
n from from each Latin index and keep the same notations for the
variables. We use $i=a-n,\;j=b-n,\;k=g-n$.

First, for the equation for the mean we simplify a relation
deduced before, (\ref{UsefulFor The Mean})

\begin{eqnarray}
  {\frac {d}{dt}}F_{{i}}=\sum _{\gamma}^{} \left( A_{{i\gamma}}-{
\it \Gamma}_{{i\gamma}} \right) F_{{\gamma}}+\sum _{k}^{} \left(
A_ {{i k}}F_{{k}}-{\it \Gamma}_{{ik}} \right) F_{{k}}\\
\nonumber =\sum _{\gamma}^{}{\cal G}_{{i\gamma}}+\sum _{k}^{}
\left( A_{{ik}}-{ \it \Gamma}_{{ik}} \right) F_{{k}}\;.
\end{eqnarray}

Note that in the sum $\sum _{\gamma}^{}{\cal G}_{{i\gamma}}$ only
one term is nonzero for $i=1\dots n$ and all terms are zero for
$i=n=1\dots 2n$. Indeed, each mRNA is controlled by only one
generator:
\begin{eqnarray}
  {\cal G}_{i\gamma}&=&\delta_{i\gamma} g_i(t)\;.
\end{eqnarray}

The above formulas tell also that only the mRNA is under the
control of the generator, not the proteins neither the DNA. To
simplify the notation we will write

\begin{eqnarray}
G_i(t)&=&g_i(t),\;\;\;i=1\dots n\;,\\
G_i(t)&=&0,\; \;\;i=n+1 \dots 2n\;.
\end{eqnarray}

 The equation for the
mean then simplifies to

\begin{center}
\fbox{
\begin{minipage}{0.5\textwidth}
\begin{eqnarray}
  {\frac {d}{dt}}<q_{{i}}>=\sum _{k}^{}
H_{{ik}} <q_{{k}}>+G_i(t)\;.
\end{eqnarray}
\end{minipage}
}
\end{center}

\subsection{Solution to the Mean and Fluctuation equations}

%\begin{eqnarray}
  %{\frac {d}{dt}}<q_{{i}}>=\sum _{k}^{}
%H_{{i,k}} <q_{{k}}>+G_i(t)\;,
%\end{eqnarray}

%\begin{eqnarray}\label{EqX}
%{\frac {d}{dt}}X_{{i,j}}=\sum
%_{k}^{}(H_{{i,k}}X_{{k,j}}+H_{{j,k}}X_
%{{k,i}})+H_{{i,j}}<q_{{j}}>+H_{{j,i}}<q_{{i}}>+2\,\delta_{{{\it
%i,j}}}\sum _{k }^{}{\it
%\Gamma}_{{j,k}}<q_{{k}}>\;,\\
%\nonumber
%\end{eqnarray}

The two equations from the previous section can now be written
using a matrix notation:

\begin{eqnarray}\label{EqMatrix}
  {\frac {d}{dt}}\mu&=&H\mu+G \;,\\
  {\frac {d}{dt}}X&=&HX+X{H}^{T}+ H {\it diag} \left( \mu \right) +{\it diag} \left(\mu \right)
 H ^{T}+2diag(\Gamma \mu)\;,
\end{eqnarray}

%with
%$\begin{eqnarray}\label{MuX}
       %\nu_i&=&<q_i>
   %\nu_{i,j}&=&\delta_{i,j}<q_i>+X_{i,j}\;,
%\end{eqnarray}

where the column vector $\mu$ has the components $\mu_i=<q_i>$,
and the matrix $X$ is related with the covariance matrix $\nu $
by: $\nu_{ij}=\delta_{ij}<q_i>+X_{ij} $, $i,j=1\dots 2n $. Here
$H^T$ is the transpose matrix of $H=A-\Gamma$ and $diag(\mu)$ has
nonzero elements only on the principal diagonal:
$diag(\mu)_{ij}=\delta_{ij}\mu_i$. We took care of the fact that
$X$ is a symmetric matrix, $X^T=X$.

The first equation in ($\ref{EqMatrix}$) has a column vector as an
unknown, $\mu $, and is easy to solve it if we use the Laplace
transform

\begin{equation}
    \mu(s)=\int_0^{\infty}e^{-st}\mu(t)dt\;.
\end{equation}
The equation for the mean becomes
\begin{equation}\label{Mu}
    s\mu(s)-\mu_0=H\mu(s)+G(s)\;,
\end{equation}

with $\mu_0$ being the value of the mean number of molecules at
time zero, when the generator was applied. Thus:

%\begin{center}
%\fbox{
%\begin{minipage}{0.45\textwidth}
\begin{equation}\label{MuSolved}
    \mu(s)=(s-H)^{-1}(G(s)+\mu_0)
\end{equation}
%\vspace{1pt}
%\end{minipage}
%}
%\end{center}

 The next goal is to solve for $X$. The second equation in ($\ref{EqMatrix}$) is a matrix equation. To find
 a solution for $X$ we transform the matrix equation into a vector
 equation. The transformation needed is (\cite{Horn} page 244):

\begin{equation}\label{vec}
   X\mapsto vec(X)\;,
\end{equation}

where the column vector $vec(X)$ contains the columns of the
matrix $X$ one on top of the next one, starting with the first
column and ending with the last column. In index notations, the
element $X_{ij}$ of the matrix $X$ gets into the line $i+m(j-1)$
in $vec(X)$ if $X$ is an $m\times m$ matrix.

The $vec$ mapping has the useful property that
\begin{eqnarray}
  vec(HX)&=&(1\otimes H) vec(X)\;,\\
  vec(XH)&=&(H^T\otimes 1) vec(X)\;,
\end{eqnarray}
were 1 is the unit matrix and $A\otimes B$ is the tensor product
of two matrices A and B. The matrix $A \otimes B$ is constructed
by substituting each element $a_{ij}$ of the matrix $A$ by the
matrix $a_{ij}B$.

The matrix equation for $X$ becomes

\begin{equation}\label{VecX}
    {\frac {d}{dt}}vec(X)=(H\otimes 1+1\otimes H)vec(X)+ \left (H\otimes 1+1\otimes H\right)vec({\it diag}\left(
    \mu\right))+2 vec(diag(\Gamma \mu))\;.
\end{equation}

The column vector $vec({\it diag}\left( \mu \right))$ can be
expressed in terms of the column vector $\mu$:

\begin{equation}
vec({\it diag}\left( \mu \right))=L\mu\;,
\end{equation}

were $L$ is a lift matrix from a space of dimension of $\mu$ to
the square of this dimension. The matrix $L$ has the block
structure

\begin{equation}
L=\left( \begin{array}{c}
  P_1 \\
  \vdots \\
  P_{2n}
\end{array} \right)\;,
\end{equation}

where $2n$ is the number of rows in $\mu$ ( n rows for mRNA and
another n for proteins). The submatrices $P_k,\;k=1...2n$ are $2n
\times 2n$ square projection matrices, with all elements zero
except one:
\begin{equation}
    (P_k)_{ab}=\delta_{ak}\delta_{bk}\;.
\end{equation}

As an example, for $n=1$ we have 1 mRNA and 1 protein and the
dimension of $L$ is $4\times 2$

\begin{equation}
   L= \left(%
\begin{array}{cc}
  1 & 0 \\
  0 & 0 \\
  0 & 0 \\
  0 & 1 \\
\end{array}%
\right)\;.
\end{equation}

With the same lift matrix $L$ we can write
\begin{equation}
    vec(diag(\Gamma \mu))=L\Gamma \mu
\end{equation}
 Denote now the Laplace
transform of $vec(X)$ as $V$. We have, from $\ref{VecX}$,

\begin{equation}
    sV(s)-V_0=(H\otimes 1+1\otimes H)V(s)+ \left(\left (H\otimes 1+1\otimes
    H\right)L+2L\Gamma\right)\mu(s)\;,
\end{equation}

%\fbox{
%\begin{minipage}{0.99\textwidth} \begin{eqnarray} \nonumber
  %V \left( s \right) &=& \left( s-{\it 1\otimes H}-{\it H\otimes 1} \right) ^{-1}
  %\left(\left( A-{{\it \Gamma}}^{{\it nd}} \right) {\it \otimes 1}+{\it 1\otimes}\, \left( A-{{\it \Gamma}}^{{\it nd}}
  %\right)\right)L(s-H)^{-1}(G(s)+\mu_0)+\\ \nonumber
%& &+\left( s-{\it 1\otimes H}-{\it H\otimes 1} \right) ^{-1}V_0
%\end{eqnarray}
%\end{minipage}
%}

%\fbox{
%\begin{minipage}{0.99\textwidth}

\begin{eqnarray}\label{FluctSol}
%\nonumber
  V \left( s \right) &=& \left( s-{\it 1\otimes H}-{\it H\otimes 1} \right) ^{-1}
  \left(\left( {\it 1\otimes H}+{\it H\otimes 1} \right) L+2\,L{\it \Gamma}\right)(s-H)^{-1}(G(s)+\mu_0)\;\;\;\;\;\;\;\;\;\;\;\;\;\\ \nonumber
& &+\left( s-{\it 1\otimes H}-{\it H\otimes 1} \right) ^{-1}V_0\;.
\end{eqnarray}
%\end{minipage}
%}

\vspace{0.5cm}
 For a diagonal $\Gamma $

%\fbox{
%\begin{minipage}{0.99\textwidth}
\begin{eqnarray}
\nonumber
  V \left( s \right) &=& \left( s-{\it 1\otimes H}-{\it H\otimes 1} \right) ^{-1}
  \left( A  {\it \otimes 1}+{\it 1\otimes}\,  A
  \right)L(s-H)^{-1}(G(s)+\mu_0)+\\ \nonumber
& &+\left( s-{\it 1\otimes H}-{\it H\otimes 1} \right) ^{-1}V_0\;,
\end{eqnarray}
%\end{minipage}
%}

with $\mu_0$ is the initial condition for the mean and $V_0$ the
initial condition for $vec(X)$.

%\fbox{
%\begin{minipage}{0.99\textwidth}

%\fbox{
%\begin{minipage}{0.98\textwidth}

From the above formula (\ref{MuSolved})  we see that the mean
values are expressed in terms of the generators through the {\it
mean transfer matrix}:

\begin{equation}\label{GreenFunctionMean}
   \frac{1}{s-H}\;.
\end{equation}

The interesting form, (\ref{FluctSol}), is the { \it fluctuation
transfer matrix}, that passes the time variation of the input
generators into the time variation of $vec(X)$:

\vspace{0.5cm}
 \fbox{
\begin{minipage}{0.98\textwidth}
\begin{eqnarray}\nonumber
%\label{GreenFunctionfFluctuation}
   \frac{1}{s-1\otimes H-H\otimes 1}\left[\left( {\it 1\otimes H}+{\it H\otimes 1} \right) L+2\,L{\it
   \Gamma}\right]\frac{1}{s-H}\nonumber\;.
\end{eqnarray}
\vspace{0.1cm}
\end{minipage}
} \vspace{0.5cm}

 For a diagonal $\Gamma$ matrix this simplifies to

\begin{equation}\label{GreenFunctionfFluctuation}
   \frac{1}{s-1\otimes H-H\otimes 1}\left[\left( {\it 1\otimes A}+{\it A\otimes 1} \right)
   L\right]\frac{1}{s-H}\;.
\end{equation}

%\end{minipage}
%}
%\end{minipage}
%}

 As an example, if  $H$ and $\Gamma$ are 2 by 2 matrices,

\begin{equation}\label{H}
   H= \left[ \begin {array}{cc} h_{{11}}&h_{{12}}\\\noalign{\medskip}h_{{21}}&h_{{22}}\end {array}
   \right]\;\;\;\; , \;\;\;\; \Gamma=\left[ \begin {array}{cc} g_{{11}}&g_{{12}}\\\noalign{\medskip}g_{{21}}&g_{{22}}\end {array} \right]
\end{equation}

we get:
\begin{equation}
    s-1\otimes H-H\otimes 1=\left[ \begin {array}{cccc} -2\,h_{{11}}+s&-h_{{12}}&-h_{{12}}&0\\\noalign{\medskip}-h_{{21}}&-h_{{11}}-h_{{22}}+s&0&-h_{{12}}
\\\noalign{\medskip}-h_{{21}}&0&-h_{{11}}-h_{{22}}+s&-h_{{12}}
\\\noalign{\medskip}0&-h_{{21}}&-h_{{21}}&-2\,h_{{22}}+s
\end {array} \right]
\end{equation}

\begin{equation}
   \left( {\it 1\otimes H}+{\it H\otimes 1} \right) L+2\,L{\it
   \Gamma} = \left[ \begin {array}{cc} 2\,h_{{11}}+2\,g_{{11}}&2\,g_{{12}}
\\\noalign{\medskip}h_{{21}}&h_{{12}}\\\noalign{\medskip}h_{{21}}&h
_{{1,2}}\\\noalign{\medskip}2\,g_{{21}}&2\,h_{{22}}+2\,g_{{22}}
\end {array} \right]
\end{equation}

\begin{eqnarray}
    \left(\left( {\it 1\otimes H}+{\it H\otimes 1} \right) L+2\,L{\it
    \Gamma}\right)\frac{1}{s-H}\;= \nonumber
\end{eqnarray}
\begin{eqnarray}
  \!\!\!\!\!\!\!\!\!\!
   = \frac{1}{\Delta}\left[ \begin {array}{cc} 2\,h_{{11}}s-2\,h_{{11}}h_{{22}}+2\,g_{{11}}s-2\,g_{{11}}h_{{22}}+2\,g_{{12}}h_{{21}}&2\,h_{{12}}h_{{11
}}+2\,h_{{12}}g_{{11}}+2\,g_{{12}}s-2\,g_{{12}}h_{{11}}
\\\noalign{\medskip}h_{{21}} \left( s-h_{{22}}+h_{{12}} \right) &h_
{{12}} \left( h_{{21}}+s-h_{{11}} \right)
\\\noalign{\medskip}h_{{2 1}} \left( s-h_{{22}}+h_{{12}}
\right) &h_{{12}} \left( h_{{21}}+ s-h_{{11}} \right)
\\\noalign{\medskip}2\,g_{{21}}s-2\,g_{{21}}h_{{
22}}+2\,h_{{21}}h_{{22}}+2\,h_{{21}}g_{{22}}&2\,g_{{21}}h_{{12}
}+2\,h_{{22}}s-2\,h_{{11}}h_{{22}}+2\,g_{{22}}s-2\,g_{{22}}h_{{1
1}}\end {array} \right]\nonumber
\end{eqnarray}

\begin{equation}
   \Delta={s}^{2}-h_{{22}}s-h_{{11}}s+h_{{11}}h_{{22}}-h_{{12}}h_{{2
1}}
\end{equation}

\subsection{An autoregulatory gene with a periodically driven
cofactor. Response of the system to an arbitrary input}

One of the most fundamental regulatory motif in a genetic network
is an autoregulatory gene through a negative feedback,
\cite{Young}. We consider the case when the gene regulation is
under the control of its own protein product and the protein
activity is modulated by a cofactor. The equation for the mean is:
\begin{equation}
{\frac {d}{dt}}\left[ \begin {array}{c} \langle r \rangle
\\\noalign{\medskip}\langle p \rangle \end {array} \right] = \left[ \begin
{array}{cc} -\gamma_r&-h\\\noalign{\medskip}k_p&-\gamma_p\end
{array} \right] \left[ \begin {array}{c} \langle r \rangle
\\\noalign{\medskip}\langle p \rangle\end {array} \right] +\left[ \begin
{array}{c} k_{{0}}+g(t)
\\\noalign{\medskip}0\end {array} \right]
\end{equation}
where the state is $q=(r,p).$ The cofactor, represented here by
the term $g(t),$ is  driven by the light generator. The cofactor
modulates the mRNA mean number $\langle r \rangle$ through an
additive coupling. For this case we use suggestive notations
$X_{11}=X_{rr}$, $X_{12}=X_{rp}$ etc.. The Laplace transform of
the quantities of interest, ($\langle r(t)\rangle,\langle
p(t)\rangle,X_{rr}(t),X_{rp}(x),X_{pp}(t)$, $g(t)$) are denoted by
the same letter but the argument being the complex frequency $s$
instead of the time $t$, like in
\begin{equation}
\langle r \rangle(s)=\int_{0}^{\infty} \langle r(t)\rangle
e^{-st}dt\;.
\end{equation}

The values of the mean number of molecules and their fluctuation,
will depend on the internal parameters $\gamma
_r,\gamma_p,h,k_p,k_0$ as well as on the external parameters of
the generator $g(t).$ Two important natural parameters of the
system play a significant role:
\begin{eqnarray}
\omega_0^2&=&hk_p+\gamma_r\gamma_p\;,
\\
\omega_1&=&\gamma_r+\gamma_p\;.
\end{eqnarray}
 The mean number of molecules are connected to the generator
 through:
\begin{equation}
\left[ \begin {array}{c} \langle r \rangle \left( s \right)
\\\noalign{\medskip}\langle p \rangle \left( s \right) \end {array} \right]
=\frac{1}{\Delta (s)}\, \left[
\begin {array} {cc}
s+\gamma_{{p}}&-h\\\noalign{\medskip}k_{{p}}&s+\gamma_{{r}}
\end {array} \right]  \left[ \begin {array}{c} g \left( s \right)
\\\noalign{\medskip}0\end {array} \right]\;,
\end{equation}
with
\begin{equation}
\Delta \left( s \right)
={s}^{2}+s\omega_{{1}}+{\omega_{{0}}}^{2}\;.
\end{equation}

The deviation from a Poisson process measured by the variable $X$
is under the generator influence also:

\begin{equation}
\nonumber \left[ \!\!\begin {array}{c}   X_{{rr}}  \left( s
\right) \\\noalign{\medskip}  X_{{rp}} \left( s \right)
\\\noalign{\medskip}  X_{{pr}}   \left( s \right)
\\\noalign{\medskip}  X_{{pp}}   \left( s \right)
\end {array} \!\!\right]\!=\!\frac{1}{\Delta_f \left( s \right)}  \left[ \begin {array}
{cc} -2\,h \left( s+2\,\gamma_{{p}} \right) k_{{p}} \left(
s+\gamma_{{ p}}-h \right) &2\,{h}^{2} \left( s+2\,\gamma_{{p}}
\right)  \left( s+k _{{p}}+\gamma_{{r}} \right)
\\\noalign{\medskip}k_{{p}} \left( s+2\, \gamma_{{p}} \right)
\left( s+2\,\gamma_{{r}} \right)  \left( s+ \gamma_{{p}}-h \right)
&-h \left( s+2\,\gamma_{{p}} \right)  \left( s+ 2\,\gamma_{{r}}
\right)  \left( s+k_{{p}}+\gamma_{{r}} \right)
\\\noalign{\medskip}k_{{p}} \left( s+2\,\gamma_{{p}} \right)  \left( s
+2\,\gamma_{{r}} \right)  \left( s+\gamma_{{p}}-h \right) &-h
\left( s +2\,\gamma_{{p}} \right)  \left( s+2\,\gamma_{{r}}
\right)  \left( s+k _{{p}}+\gamma_{{r}} \right)
\\\noalign{\medskip}2\,{k_{{p}}}^{2}
 \left( s+2\,\gamma_{{r}} \right)  \left( s+\gamma_{{p}}-h \right) &-2
\,h \left( s+2\,\gamma_{{r}} \right) k_{{p}} \left(
s+k_{{p}}+\gamma_{ {r}} \right) \end {array} \right]  \left[\!\!
\begin {array}{c} g \left( s
 \right) \\\noalign{\medskip}0\end {array} \!\!\right],
 \nonumber
\end{equation}
with
\begin{equation}
\Delta_f \left( s \right) = \left( s+\omega_{{1}} \right)\left(
{s}^{2}+s \omega_{{1}}+{\omega_{{0}}}^{2} \right) \left( {s}^{2
}+2\,s\omega_{{1}}+4\,{\omega_{{0}}}^{2} \right)
\end{equation}

\subsection{The step and the periodic stimuli}

There are two cases of interest to us, a step stimulus and a
periodic one.

For a step stimulus:
\begin{equation}
G(s)=\frac{G}{s}\;.
\end{equation}
We consider that the system is in a steady state before we apply
the step stimulus. The steady state is governed by the translation
rate $k_0$. For a stable system ( Re($\lambda _{1,2})>0$), the
mean number of molecules decay exponentially to zero.
\begin{eqnarray}
\label{ImpuseResponse}
 \langle r\left( t \right)\rangle ={\frac { \gamma_{{p}}}{
\lambda_{{1}}\lambda_{{2}}}}\,k_{{0}}+{\frac { \gamma_{{p}}}{
\lambda_{{1}}\lambda_{{2}}}}\,G+{\frac { \left(
\lambda_{{1}}-\gamma_{{p} } \right) }{\lambda_{{1}} \left(
\lambda_{{2}}- \lambda_{{1}} \right)
}}G\,{e^{-\lambda_{{1}}t}}+{\frac {\left(
\lambda_{{2}}-\gamma_{{p}}
 \right) }{\lambda_{{2}} \left( \lambda_{{1}}-
\lambda_{{2}} \right) }}G\,{e^{-\lambda_{{2}}t}}
\\
\langle p \left( t \right)\rangle ={\frac {   k_{{p}}}{
\lambda_{{1}}\lambda_{{2}}}}\,k_{{0}}+{\frac
{k_{{p}}}{\lambda_{{2}}\lambda_{{1}}}}\,G+{\frac
{k_{{p}}}{\lambda_{{1 }} \left( \lambda_{{1}}-\lambda_{{2}}
\right) }}\,G{e^{-\lambda_{{1}}t}}+{\frac {k_{{p}}}{\lambda_{{2}}
\left( \lambda_{{2}}-\lambda_{{1}}
 \right) }}G{e^{
-\lambda_{{2}}t}}
%\\
%r(t)={\frac { \left( -\gamma_{{p}}+\lambda_{{1}} \right)
%{e^{-\lambda_{{1}} t}}+ \left( \gamma_{{p}}-\lambda_{{2}} \right)
%{e^{-\lambda_{{2}}t}}}{ -\lambda_{{2}}+\lambda_{{1}}}}r(0)+ {\frac
%{h{e^{-\lambda_{{1}}t}}-h{e^{-\lambda_{{2}}t}}}{-\lambda_{{2}}+
%\lambda_{{1}}}}p(0)
%\\
%p(t)={\frac{-k_{{p}}{e^{-\lambda_{{1}}t}}+k_{{p}}{e^{-\lambda_{{2}}t}}}{-
%\lambda_{{2}}+\lambda_{{1}}}}r(0)+ {\frac { \left(
%-\gamma_{{r}}+\lambda_{{1}} \right) {e^{-\lambda_{{1}} t}}+ \left(
%\gamma_{{r}}-\lambda_{{2}} \right) {e^{-\lambda_{{2}}t}}}{
%-\lambda_{{2}}+\lambda_{{1}}}}p(0)
\end{eqnarray}

where $\lambda_1$ and $\lambda_2$ are the eigenvalues of $H$:
$\Delta \left( s \right) = \left( s+\lambda_{{1}} \right)  \left(
s+ \lambda_{{2}} \right)\;, $
\begin{eqnarray}
\label{lambda1lambda2}
 \lambda_{{1}}&=&1/
2\,\omega_{{1}}-1/2\,\sqrt
{{\omega_{{1}}}^{2}-4\,{\omega_{{0}}}^{2}}\;,
\\
\lambda_{{2}}&=&1 /2\,\omega_{{1}}+1/2\,\sqrt
{{\omega_{{1}}}^{2}-4\,{\omega_{{0}}}^{2}}\;.
\end{eqnarray}
For fluctuations we get also exponentially decaying responses to
the step stimulus:
\begin{eqnarray}\label{FlucImplulseResponse}
%\nonumber
 X_{{{\it rr}}} \left( t \right) \!\! &=&\!
\!X_{{{\it rr},0}}+\!X_{{{\it rr},\omega_
{{1}}}}{e^{-\omega_{{1}}t}}+\!X_{{{\it
rr},\lambda_{{1}}}}{e^{-\lambda_{ {1}}t}}+\!X_{{{\it
rr},\lambda_{{2}}}}{e^{-\lambda_{{2}}t}}+\!X_{{{\it rr}
,2\,\lambda_{{1}}}}{e^{-2\,\lambda_{{1}}t}}+\!X_{{{\it
rr},2\,\lambda_{{ 2}}}}{e^{-2\,\lambda_{{2}}t}} \;,\;\;\;\;
\\ \nonumber
X_{{{\it rp}}} \left( t \right) \!\! &=&\!\!X_{{{\it
rp},0}}+\!X_{{{\it rp},\omega_
{{1}}}}{e^{-\omega_{{1}}t}}+\!X_{{{\it
rp},\lambda_{{1}}}}{e^{-\lambda_{ {1}}t}}+\!X_{{{\it
rp},\lambda_{{2}}}}{e^{-\lambda_{{2}}t}}+\!X_{{{\it rp}
,2\,\lambda_{{1}}}}{e^{-2\,\lambda_{{1}}t}}+\!X_{{{\it
rp},2\,\lambda_{{ 2}}}}{e^{-2\,\lambda_{{2}}t}}\;,
\\ \nonumber
X_{{{\it pp}}} \left( t \right) \!\! &=&\!\!X_{{{\it
pp},0}}+\!X_{{{\it pp},\omega_
{{1}}}}{e^{-\omega_{{1}}t}}+\!X_{{{\it
pp},\lambda_{{1}}}}{e^{-\lambda_{ {1}}t}}+\!X_{{{\it
pp},\lambda_{{2}}}}{e^{-\lambda_{{2}}t}}+\!X_{{{\it pp}
,2\,\lambda_{{1}}}}{e^{-2\,\lambda_{{1}}t}}+\!X_{{{\it
pp},2\,\lambda_{{ 2}}}}{e^{-2\,\lambda_{{2}}t}}\;.
\end{eqnarray}
The coefficients from the above formulas are collected in the
following matrices

\begin{equation}
\left[ \begin {array}{c} X_{{{\it
rr},0}}\\\noalign{\medskip}X_{{{\it
rr},\omega_{{1}}}}\\\noalign{\medskip}X_{{{\it rr},\lambda_{{1}}}}
\\\noalign{\medskip}X_{{{\it rr},\lambda_{{2}}}}\\\noalign{\medskip}X_
{{{\it rr},2\,\lambda_{{1}}}}\\\noalign{\medskip}X_{{{\it rr},2\,
\lambda_{{2}}}}\end {array} \right] = \left[ \begin {array}{c}
{\frac {hk_{{p}}G \left( h-\gamma_{{p}} \right)
\gamma_{{p}}}{{\lambda_{{2}}}
^{2}{\lambda_{{1}}}^{2}\omega_{{1}}}}+{\frac {hk_{{p}}k_{{0}}
\left( h -\gamma_{{p}} \right)
\gamma_{{p}}}{{\lambda_{{2}}}^{2}{\lambda_{{1}}}
^{2}\omega_{{1}}}}\\\noalign{\medskip}2\,{\frac {hk_{{p}}G \left(
-2\, \gamma_{{p}}+\omega_{{1}} \right)  \left(
-\gamma_{{p}}+h+\omega_{{1}}
 \right) }{ \left( -2\,\lambda_{{1}}+\omega_{{1}} \right)  \left( -
\lambda_{{2}}+\omega_{{1}} \right)  \left(
\omega_{{1}}-2\,\lambda_{{2 }} \right)  \left(
-\lambda_{{1}}+\omega_{{1}} \right) \omega_{{1}}}}
\\\noalign{\medskip}-2\,{\frac {hk_{{p}}G \left( -\lambda_{{1}}+2\,
\gamma_{{p}} \right)  \left( \lambda_{{1}}+h-\gamma_{{p}} \right)
}{
 \left( \lambda_{{1}}-2\,\lambda_{{2}} \right)  \left( -\lambda_{{2}}+
\lambda_{{1}} \right) {\lambda_{{1}}}^{2} \left(
-\lambda_{{1}}+\omega _{{1}} \right)
}}\\\noalign{\medskip}-2\,{\frac {hk_{{p}}G \left( 2\,
\gamma_{{p}}-\lambda_{{2}} \right)  \left(
\lambda_{{2}}+h-\gamma_{{p} } \right) }{{\lambda_{{2}}}^{2} \left(
-\lambda_{{2}}+\lambda_{{1}}
 \right)  \left( -\lambda_{{2}}+\omega_{{1}} \right)  \left( 2\,
\lambda_{{1}}-\lambda_{{2}} \right) }}\\\noalign{\medskip}{\frac
{hk_{ {p}}G \left( -\lambda_{{1}}+\gamma_{{p}} \right)  \left(
2\,\lambda_{{ 1}}+h-\gamma_{{p}} \right) }{{\lambda_{{1}}}^{2}
\left( -\lambda_{{2}} +\lambda_{{1}} \right)  \left(
2\,\lambda_{{1}}-\lambda_{{2}} \right)
 \left( -2\,\lambda_{{1}}+\omega_{{1}} \right) }}\\\noalign{\medskip}{
\frac {hk_{{p}}G \left( \gamma_{{p}}-\lambda_{{2}} \right)  \left(
2\, \lambda_{{2}}+h-\gamma_{{p}} \right) }{{\lambda_{{2}}}^{2}
\left( - \lambda_{{2}}+\lambda_{{1}} \right)  \left(
\lambda_{{1}}-2\,\lambda_{ {2}} \right)  \left(
\omega_{{1}}-2\,\lambda_{{2}} \right) }}
\end {array} \right]\;,
\end{equation}

\begin{equation}
\left[ \begin {array}{c} X_{{{\it
rp},0}}\\\noalign{\medskip}X_{{{\it
rp},\omega_{{1}}}}\\\noalign{\medskip}X_{{{\it rp},\lambda_{{1}}}}
\\\noalign{\medskip}X_{{{\it rp},\lambda_{{2}}}}\\\noalign{\medskip}X_
{{{\it rp},2\,\lambda_{{1}}}}\\\noalign{\medskip}X_{{{\it rp},2\,
\lambda_{{2}}}}\end {array} \right] = \left[ \begin {array}{c} -{
\frac {k_{{p}} \left( k_{{0}}+G \right) \gamma_{{r}}\gamma_{{p}}
 \left( h-\gamma_{{p}} \right) }{{\lambda_{{2}}}^{2}{\lambda_{{1}}}^{2
}\omega_{{1}}}}\\\noalign{\medskip}{\frac {k_{{p}}G \left(
\omega_{{1} }-2\,\gamma_{{r}} \right)  \left(
-2\,\gamma_{{p}}+\omega_{{1}}
 \right)  \left( -\gamma_{{p}}+h+\omega_{{1}} \right) }{\omega_{{1}}
 \left( \omega_{{1}}-2\,\lambda_{{2}} \right)  \left( -\lambda_{{2}}+
\omega_{{1}} \right)  \left( -\lambda_{{1}}+\omega_{{1}} \right)
 \left( -2\,\lambda_{{1}}+\omega_{{1}} \right) }}\\\noalign{\medskip}{
\frac {k_{{p}}G \left( -\lambda_{{1}}+2\,\gamma_{{r}} \right)
\left( -\lambda_{{1}}+2\,\gamma_{{p}} \right)  \left(
\lambda_{{1}}+h-\gamma_ {{p}} \right) }{{\lambda_{{1}}}^{2} \left(
-\lambda_{{2}}+\lambda_{{1} } \right)  \left(
-\lambda_{{1}}+\omega_{{1}} \right)  \left( \lambda_
{{1}}-2\,\lambda_{{2}} \right) }}\\\noalign{\medskip}{\frac
{k_{{p}}G
 \left( -\lambda_{{2}}+2\,\gamma_{{r}} \right)  \left( 2\,\gamma_{{p}}
-\lambda_{{2}} \right)  \left( \lambda_{{2}}+h-\gamma_{{p}}
\right) }{
 \left( -\lambda_{{2}}+\omega_{{1}} \right)  \left( 2\,\lambda_{{1}}-
\lambda_{{2}} \right)  \left( -\lambda_{{2}}+\lambda_{{1}} \right)
{ \lambda_{{2}}}^{2}}}\\\noalign{\medskip}-{\frac {k_{{p}}G \left(
- \lambda_{{1}}+\gamma_{{r}} \right)  \left(
-\lambda_{{1}}+\gamma_{{p}}
 \right)  \left( 2\,\lambda_{{1}}+h-\gamma_{{p}} \right) }{{\lambda_{{
1}}}^{2} \left( -\lambda_{{2}}+\lambda_{{1}} \right)  \left( 2\,
\lambda_{{1}}-\lambda_{{2}} \right)  \left(
-2\,\lambda_{{1}}+\omega_{ {1}} \right)
}}\\\noalign{\medskip}-{\frac {k_{{p}}G \left( -\lambda_{
{2}}+\gamma_{{r}} \right)  \left( \gamma_{{p}}-\lambda_{{2}}
\right)
 \left( 2\,\lambda_{{2}}+h-\gamma_{{p}} \right) }{ \left( -\lambda_{{2
}}+\lambda_{{1}} \right)  \left( \lambda_{{1}}-2\,\lambda_{{2}}
 \right)  \left( \omega_{{1}}-2\,\lambda_{{2}} \right) {\lambda_{{2}}}
^{2}}}\end {array} \right]\;,
\end{equation}

\begin{equation}
\left[ \begin {array}{c} X_{{{\it
pp},0}}\\\noalign{\medskip}X_{{{\it
pp},\omega_{{1}}}}\\\noalign{\medskip}X_{{{\it pp},\lambda_{{1}}}}
\\\noalign{\medskip}X_{{{\it pp},\lambda_{{2}}}}\\\noalign{\medskip}X_
{{{\it pp},2\,\lambda_{{1}}}}\\\noalign{\medskip}X_{{{\it pp},2\,
\lambda_{{2}}}}\end {array} \right] = \left[ \begin {array}{c} -{
\frac {{k_{{p}}}^{2} \left( k_{{0}}+G \right) \gamma_{{r}} \left(
h- \gamma_{{p}} \right)
}{{\lambda_{{2}}}^{2}{\lambda_{{1}}}^{2}\omega_{{
1}}}}\\\noalign{\medskip}-2\,{\frac {{k_{{p}}}^{2}G \left(
-\gamma_{{p }}+h+\omega_{{1}} \right)  \left(
\omega_{{1}}-2\,\gamma_{{r}}
 \right) }{\omega_{{1}} \left( \omega_{{1}}-2\,\lambda_{{2}} \right)
 \left( -\lambda_{{2}}+\omega_{{1}} \right)  \left( -\lambda_{{1}}+
\omega_{{1}} \right)  \left( -2\,\lambda_{{1}}+\omega_{{1}}
\right) }}
\\\noalign{\medskip}2\,{\frac {{k_{{p}}}^{2}G \left( \lambda_{{1}}+h-
\gamma_{{p}} \right)  \left( -\lambda_{{1}}+2\,\gamma_{{r}}
\right) }{ {\lambda_{{1}}}^{2} \left( -\lambda_{{2}}+\lambda_{{1}}
\right)
 \left( \lambda_{{1}}-2\,\lambda_{{2}} \right)  \left( -\lambda_{{1}}+
\omega_{{1}} \right) }}\\\noalign{\medskip}2\,{\frac
{{k_{{p}}}^{2}G
 \left( \lambda_{{2}}+h-\gamma_{{p}} \right)  \left( -\lambda_{{2}}+2
\,\gamma_{{r}} \right) }{ \left( -\lambda_{{2}}+\lambda_{{1}}
\right)
 \left( 2\,\lambda_{{1}}-\lambda_{{2}} \right)  \left( -\lambda_{{2}}+
\omega_{{1}} \right)
{\lambda_{{2}}}^{2}}}\\\noalign{\medskip}-{\frac {{k_{{p}}}^{2}G
\left( 2\,\lambda_{{1}}+h-\gamma_{{p}} \right)
 \left( -\lambda_{{1}}+\gamma_{{r}} \right) }{{\lambda_{{1}}}^{2}
 \left( 2\,\lambda_{{1}}-\lambda_{{2}} \right)  \left( -\lambda_{{2}}+
\lambda_{{1}} \right)  \left( -2\,\lambda_{{1}}+\omega_{{1}}
\right) } }\\\noalign{\medskip}-{\frac {{k_{{p}}}^{2}G \left(
2\,\lambda_{{2}}+h -\gamma_{{p}} \right)  \left(
-\lambda_{{2}}+\gamma_{{r}} \right) }{{ \lambda_{{2}}}^{2} \left(
\lambda_{{1}}-2\,\lambda_{{2}} \right)
 \left( -\lambda_{{2}}+\lambda_{{1}} \right)  \left( \omega_{{1}}-2\,
\lambda_{{2}} \right) }}\end {array} \right]\;.
\end{equation}

 For the periodic case with an input frequency $\omega $ and
amplitude $a$, $g(t)=k_0+acos(\omega t)$, ( $k_0$ is a baseline
not controlled by the exterior light input)
\begin{equation}
g(s)=\frac{k_0}{s}+\frac{a}{s^2+\omega^2}\;.
\end{equation}

We keep only the stationary solutions in the response ( in
practice we wait for the transients to become small enough)
\begin{eqnarray}\label{HydroPump}
\langle r \left( t \right)\rangle
&=&R_{{0}}+R_{{1}}{e^{i\omega\,t}}+ { R}_{{1}}^{*}{e^{-i
\omega\,t}}\;,
\\
\langle p \left( t \right)\rangle
&=&P_{{0}}+P_{{1}}{e^{i\omega\,t}}+{P}_{{1}}^{*}{e^{-i
\omega\,t}}\;,
\\
X_{{rr}} \left( t \right)
&=&X_{{r,0}}+X_{{r,1}}{e^{i\omega\,t}}+{X}^{*}_{{r,1}
}{e^{-i\omega\,t}}\;,
\\
X_{{rp}} \left( t \right)
&=&X_{{rp,0}}+X_{{rp,1}}{e^{i\omega\,t}}+{X}^{*}_{{rp,1}
}{e^{-i\omega\,t}}\;,
\\
X_{{pr}} \left( t \right)
&=&X_{{p,0}}+X_{{p,1}}{e^{i\omega\,t}}+{X}^{*}_{{p,1}
}{e^{-i\omega\,t}}\;.
\end{eqnarray}
The star $*$ means complex conjugation. In terms of the parameters
that constitutes the autoregulatory system we have:

\begin{eqnarray}
\label{PeriodicResponse}
 R_{{0}}&=&{\frac
{\gamma_{{p}}k_{{0}}}{\gamma_{{r}}\gamma_{{p}}+hk_{{p}} }}\;,
\\
P_{{0}}&=&{\frac
{k_{{p}}k_{{0}}}{\gamma_{{r}}\gamma_{{p}}+hk_{{p}} }}\;,
\\
R_{{1}}&=&1/2\,{\frac {a \left( \gamma_{{p}}+i\omega \right)
}{{\omega_{{0}}}^{2 }-{\omega}^{2}+i\omega\,{\it \omega_1}}}\;,
\\
P_{{1}}&=&1/2\,{\frac {{\it k_p}\,a}{-{\omega}^{2}+{{\it
\omega_0}}^{2}+i\omega\,{ \it \omega_1}}}\;,
\\
X_{{r,0}}&=&{\frac {k_{{0}} \left( h-\gamma_{{p}} \right)
hk_{{p}}\gamma _{{p}}}{{\omega_{{0}}}^{4}\omega_{{1}}}}\;,
\\
X_{{{\it rp},0}}&=&{\frac {k_{{0}}\gamma_{{r}}\gamma_{{p}}k_{{p}}
 \left( \gamma_{{p}}-h \right)
 }{\omega_{{1}}{\omega_{{0}}}^{4}}}\;,
\\
X_{{p,0}}&=&{\frac {{k_{{p}}}^{2}k_{{0}} \left( \gamma_{{p}}-h
\right) \gamma_{{r}}}{{\omega_{{0}}}^{4}\omega_{{1}}}}\;,
\\
X_{{r,1}}&=&{\frac {-ia \left( -i\gamma_{{p}}+\omega+ih \right)
\left( -\omega+2\,i\gamma_{{p}} \right) hk_{{p}}}{ \left(
-{\omega}^{2}+{ \omega_{{0}}}^{2}+i\omega\,\omega_{{1}} \right)
\left( -{\omega}^{2}+
2\,i\omega\,\omega_{{1}}+4\,{\omega_{{0}}}^{2} \right)  \left(
-\omega +i\omega_{{1}} \right) }}\;,
\\
X_{{{\it rp},1}}&=&-1/2\,{\frac {ak_{{p}} \left(
\omega-i\gamma_{{p}}+ih
 \right)  \left( \omega-2\,i\gamma_{{r}} \right)  \left( \omega-2\,i
\gamma_{{p}} \right) }{ \left(
{\omega}^{2}-{\omega_{{0}}}^{2}-i\omega \,\omega_{{1}} \right)
\left( {\omega}^{2}-4\,{\omega_{{0}}}^{2}-2\,i
\omega\,\omega_{{1}} \right)  \left( \omega-i\omega_{{1}} \right)
}}\;,
\\
X_{{p,1}}&=&{\frac {ia \left( -i\gamma_{{p}}+\omega+ih \right)
\left( \omega-2\,i \gamma_{{r}} \right) {k_{{p}}}^{2}}{ \left(
{\omega}^{2}-{\omega_{{0}} }^{2}-i\omega\,\omega_{{1}} \right)
\left( {\omega}^{2}-2\,i\omega\,
\omega_{{1}}-4\,{\omega_{{0}}}^{2} \right)  \left(
\omega-i\omega_{{1} } \right) }}\;.
\end{eqnarray}

The impact of the natural (internal) frequencies $\omega _0$ and
$\omega _1$ on the protein and mRN levels and fluctuation can be
read out from the absolute values of the denominators of the mean
and X:
\begin{eqnarray}
\Delta&=&|det(i\omega-H)|^2={\omega_{{0}}}^{2} \left(  \left(
{\omega}^{2}-{\omega_{{0}}}^{ 2} \right)
^{2}+{\omega}^{2}{\omega_{{1}}}^{2} \right)\;,
\\
\Delta_{{f}}&=&|det(i\omega-1\otimes H-H\otimes
1)|^2=4\,{\omega_{{1}}}^{2}{\omega_{{0}}}^{2} \left( {\omega_{{
1}}}^{2}+{\omega}^{2} \right) ^{2} \left( \left( {\omega}^{2}-4\,{
\omega_{{0}}}^{2} \right) ^{2}+4\,{\omega}^{2}{\omega_{{1}}}^{2}
 \right).\;\;\;\;
\end{eqnarray}
We observe that $\omega_0$ is a resonance for the mean and X),
whereas $2\omega_0$ is only for X.

 Beside the ratios expressed by formulas (5) and (6) from the main paper, we can form different combinations between the periodic response variables that become useful for estimating the order of
magnitude of the coefficients $k,h,\gamma _r, \gamma _p$ (we
consider the case when no experimental noise is present):

\begin{eqnarray}
{\frac {   \left| R_{{1}} \right|   ^{2}}{
 \left| P_{{1}} \right|  ^{2}}}&=&{\frac {1}{{k_{{p}}
}^{2}}}\,{\omega}^{2}+{\frac {{\gamma_{{p}}}^{2}}{k_{{p}}^2}}\;,
\\
{\frac {X_{{r,0}}}{X_{{p,0}}}}&=&-{\frac
{h\gamma_{{p}}}{k_{{p}}\gamma_{ {r}}}}\;,
\\
R_{{1,\omega=0}}&=&\frac{1}{2}\,{\frac
{\gamma_{{p}}}{{\omega_{{0}}}^{2}}}\,a\;.
\end{eqnarray}
From the first relation we can estimate $k_p$ and $\gamma_p$. From
the second one we estimate the ratio $\frac {h}{\gamma_r}.$ The
third equation gives the variation of mRNA amplitude with the
input amplitude $a$ for small $\omega\,.$ From this relation we
estimate $\omega_0\,.$ The mRNA degradation coefficient $\gamma_r$
can now be obtained from
\begin{equation}
\gamma_r=\omega_0^2/(\gamma_p+\frac {h}{\gamma_r}\,k_p)\;.
\end{equation}
Now we have $h$ from $h/\gamma_r\,.$ The last parameter $k_0$
comes from
\begin{equation}
R_{{0}}={\frac {\gamma_{{p}}k_{{0}}}{{\omega_{{0}}}^{2}}}\;.
\end{equation}
There are other interesting ratios worth to be written down:
\begin{eqnarray}
P_{{1,\omega=0}}&=&\frac{1}{2}\,{\frac
{k_{{p}}}{{\omega_{{0}}}^{2}}}\,a\;,
\\
{\frac {   \left| X_{{p,1}} \right|   ^{2}}{
 \left| X_{{r,1}} \right|   ^{2}}}&=&{\frac {{h}^{2} \left( {
\omega}^{2}+4\,{\gamma_{{p}}}^{2} \right) }{{k_{{p}}}^{2} \left( {
\omega}^{2}+4\,{\gamma_{{r}}}^{2} \right) }}\;,
\\
{\frac {R_{{0}}}{P_{{0}}}}&=&{\frac {\gamma_{{p}}}{k_{{p}}}}\;.
\end{eqnarray}
These relations can be used to further verify the validity of the
model, once we estimated the parameters.

\subsection{Fluctuation resonance}

We want to find a driving frequency for which the fluctuations
dominates over the mean values. For such a frequency the system
will be in a pure fluctuation resonance. In such a situation the
molecular noise can drive the cell out of its equilibrium state,
which can have dramatic consequence on the cell fate. At the
fluctuation resonance frequency, the deviation from a Poissonian
process, measured by the quantity $X$, should be very high. To
measure this deviation we consider the ratio of the fluctuation
amplitude $\mid X_{p1}\mid$ over the mean amplitude $\mid P_1\mid$
(an analog of the Fano factor in frequency domain):

\begin{equation}\label{RatioResonance}
    \frac{\mid X_{p1}\mid}{\mid P_1\mid}=\left(4\,{k_{{p}}}^{2}{\frac { \left( {\omega}^{2}+ \left( h-{\gamma_p} \right) ^{2} \right)
 \left( {\omega}^{2}+4\,{\gamma_{{r}}}^{2} \right) }{ \left(  \left( {
\omega}^{2}-4\,{\omega_{{0}}}^{2} \right)
^{2}+4\,{\omega}^{2}{\omega_ {{1}}}^{2} \right)  \left(
{\omega}^{2}+{\omega_{{1}}}^{2} \right) }}\right)^{1/2}\;.
\end{equation}

For systems for which $\omega_0\gg\omega_1$ we can se a resonance
for fluctuations but not for the mean values at the input
frequency $\omega =2\omega_0.$ A plot of this ratio is presented
in Fig.3. We notice that the width and the hight of the resonance
are inverse proportional. The parameters for which we see the
resonance in Fig.3 doesn't belong to case we studied for step
stimulus ($\lambda_{1,2}$ are real numbers there and complex
here). The response to the step stimulus for systems that can
enter into fluctuation resonance is a superposition of damped
oscillations. Even in this situation the transients are gone after
few periods.

\subsection{The Genetic Network Spectral Function}

The time response (mean and fluctuation) of the autoregulatory
system to a step stimulus can be expressed in general as a sum of
6 terms
\begin {equation}
\label{f}
 f_{exp}(t)=S_{exp,0}+S_{exp,1}e^{-\eta _1 t}+S_{exp,2}e^{-\eta _2 t}+S_{exp,_3}e^{-\eta _3
t}+S_{exp,4}e^{-\eta _4 t}+S_{exp,5}e^{-\eta _5 t}\;.
\end{equation}
Only three of these terms are present in the mean. For the purpose
of the following analysis, we will consider only the case when all
$\eta 's$ are positive, which is equivalent with
$\omega_1>2\omega_0.$
 The asymptotic response, of the same autoregulatory system, to a periodic stimulus has the
form
\begin{equation}
\label{fp}
 f_{per} \left( t \right)
=S_{{per,0}}+S_{{per,1}}{e^{i\omega\,t}}+{ {
S}^{*}_{{per,1}}}{e^{- i\omega\,t}}\;,
\end{equation}
for both the mean and the fluctuation. The parameters of the
system $k_p,h,\gamma _r,\gamma_p$ are hidden in the coefficients
$S_{exp,i}$ or $S_{per,j}$, $i=0,\dots ,5$, $j=0,1$. For more
complex genetic network, the time evolution of the measured
quantity $f(t)$ can be expressed as
\begin{equation}
\label{generalproblem} f \left( t \right) =\int _{x_1}^{x_2}\!S
\left( x \right) K \left( x\, t \right) {dx}\;.
\end{equation}

Here $S(x)$ is the {\it spectral function} that contains the
information about the genetic network and  $K(xt)$ is the kernel
that depends only on the type of the stimulus ( i.e. on the
experimental design). Indeed, for an autoregulatory network, using
the Dirac's $\delta$-function, we have

\begin{eqnarray}\label{int}
f_{exp} \left( t \right) &=&\int _{a}^{b }\!S_{exp} \left( x
\right)
{e^{-x \,t}}{dx}\;,\\
 S_{exp} \left( x \right) &=&\sum
_{i=1}^{6}S_{{exp,i}}\delta
\left( x-\eta_{{i} } \right)\;,\\
 K_{exp}(xt)&=&e^{-xt}.
\end{eqnarray}

The values $a$ and $b$ are chosen such that the spectrum
$S_{exp}(x)$ is zero outside the interval $[a,b]$.

 For the periodic stimulus, we have a similar representation for the
spectral function $S(x)$ but the kernel is different

\begin{eqnarray}\label{intp}
f_{per} \left( t \right) &=&\int _{-\Omega }^{\Omega }\!S_{per}
\left( x
 \right) {e^{ix\,t}}{dx}\;,\\
 S_{per} \left( x \right) &=&S_{{per,0}}\delta
\left( x \right)+S_{{per,1}}\delta \left( x -\omega\right)+{
S^{*}}_{{per,1}}\delta
\left( x +\omega\right)\;,\\
 K_{per}(xt)&=&e^{-ixt}.
\end{eqnarray}

with $\Omega >\omega .$

The topology of the genetic network is reflected in the spectral
function $S(x)$. Given a set of measured data, first we have to
recover the spectral function of the network and then from it the
parameters of the network. If we lack a good model for the
topology of the genetic network we cannot find the parameters of
the network, but we can recover the spectral function $S(x)$ from
the data (the kernel $K(xt)$ does not depend on the network). Thus
different genetic networks can be compared using their spectral
functions. However, the spectral function depends on the
experimental design. We proved for  the autoregulatory system that
the spectral function $S_{per}$ is much simpler than $S_{exp}$. We
want to show that there is even a deeper difference between these
two experimental designs. Namely, in the presence of experimental
noise, it is much easier to recover $S_{per}$ from the
experimental data than $S_{exp}$. This phenomena appeared in other
branches of science and in many different forms. To adapt it to
biology, we noticed that a legitimate question from a molecular
biologist is: instead of creating new assays to measure $S_{per}$
why is not enough to increase the number of replicates to obtain
an accurate $S_{exp}$ ? We will prove that the number of
replicates for $S_{exp}$ growth exponentially with the accuracy.
In what follows we collect and use for our specific problem,
results form \cite{Italy},\cite{Slepian}.

In laboratory measurements, we don't have $f(t)$ for all values of
$t$. Rather, we have samples of it at discrete time points. For
the periodic stimulus, we measure $f(\tau n)$, where
$n=0,1,\dots,N $. As a working example, consider the samples of
the mean of the mRNA, $r(n)\equiv \langle r(\tau
n)\rangle,n=0\dots N-1,\;$. The unknown spectrum $S_{per}(\omega)$
and $r(n)$ are related through the equation:

\begin{equation}
r \left( n\right) =\int
_{-\Omega}^{\Omega}\!{e^{in\tau\,\omega}}S_{per} \left( \omega
\right) {d\omega}\;.
\end{equation}

 There are three parameters in the problem: $\tau,\Omega, N.$
The sampling parameter $\tau$ must be such that the input
frequency $\omega_{in}$ can be detected in the output data, that
is $\tau \leq \pi /\omega_{in}$. The  frequency $\Omega $ should
be greater than the input frequency $\omega_{in}$. There is no
condition on the number of points $N.$ Because we have a finite
number $N$ of measured data points, the spectrum $S_{per}(\omega
)$ can only be approximated  as a weighted sum of $N$ functions
$\Phi_k(\omega)$ ( see Appendix 1 and \cite{Slepian})

\begin{equation}
\label{SolutionSpectrumWithoutError}
 {\tilde
S}_{per}\left( \omega \right) =\sum
_{k=0}^{N-1}{s_k}\Phi_k(\omega).
\end{equation}
The functions  $\Phi_k(\omega), k=0\cdots N-1$ come from a
eigenvalue problem for an $N\times N$ matrix (see Appendix 1 at
the end of this Supporting Material). Now,the experimental noise
will alter the coefficients $s_k$ so the recovered spectrum will
be:

\begin{equation}
\label{SSE}
 {\tilde
S}_{per}\left( \omega \right) =\sum _{k=0}^{N-1}{\left(s_k+\frac
{\epsilon_k}{ \beta_{{k}}}\right)}\Phi_k(\omega)\;,
\end{equation}

where the $\epsilon_k$ are the noise coefficients. The numbers
$\beta _k,\;k=1\cdots N-1,$ come from the same eigenvalue problem
as before and they depend only on the parameters $\tau ,\Omega, N$
and not on the noise coefficients $\epsilon_k.$
 Due to noise, we cannot use all $N$ terms in (\ref{SSE}), but only the first $J_{p}$, for
 which
 \begin{equation}
 \label{SignalToNoiseConditionPeriodic1}
 \frac{1}{\beta_k}<\frac{s_k}{\epsilon_k},\;\;\;\;\;\;\;k=1,\cdots,J_{p}\;.
 \end{equation}

 The right hand side of (\ref{SignalToNoiseConditionPeriodic1}) is the Signal to Noise Ration (SNR)
 and for simplicity we will consider that is independent of the
 index $k.$ The numbers $\beta_k$ decrease as $k$ increase and so
 the condition for the cutoff $J_{p}$ is simple

\begin{equation}
 \label{SignalToNoiseConditionPeriodic2}
 \frac{1}{\beta_{J_{p}}}<SNR<\frac{1}{\beta_{(J_{p}+1)}}\;.
 \end{equation}

 The exponential case can be developed parallel to the periodic
case,\cite{Italy}. The problem now reads like
\begin{equation}
r \left( n \right) =\int _{a}^{b}\!{e^{-p_{{n}}\lambda}}S_{exp}
\left( \lambda \right) {d\lambda}\;.
\end{equation}

 Unlike for the periodic case, here a geometric sampling is optimum
 \cite{Italy}

\begin{equation}
p_n=\frac{q}{a}\Delta^n,\;\; n=1\dots N\;.
\end{equation}
The limits $a$ and $b$ are chosen so that the spectrum is nonzero
only  inside $[a,b]$. For the periodic case we know the input
frequency so we don't have to guess an interval $[a,b]$ as we have
to do for the exponential case. Only the ratio $\gamma=b/a$ is
important, as we see by  changing the variable $\lambda =ax$
\begin{equation}
r \left( n \right) =\int _{1}^{\gamma}\!{e^{-ap_{{n}}x}}S_{exp}
\left( ax \right) a{dx}\;.
\end{equation}
Similar to the periodic case, solving an eigenvalue problem we can
find an N-dimensional approximation to the spectrum. Because of
the experimental noise we can use only $J_e$ degrees of freedom,
not $N:$

\begin{equation}
\label{SolExpError} {\tilde S}_{exp} \left( \lambda \right) = \sum
_{k=1}^{J_{e}}{\left(s_k+\frac {\epsilon_k}{\alpha_{{k }}}\right)}
\Psi_{{k}} \left( {\lambda} \right)\;.
\end{equation}
Here the terms $\epsilon_k$ are due to random experimental errors.
Again, the functions $\Psi_k(\lambda)$ and the numbers $\alpha_k$
come from an eigenvalue problem ( different from the periodic one)
and they don't depend on the noise but only on the parameters
$a,q,\Delta,\gamma,N$ (actually, the numbers $\alpha _k$ do not
depend on the parameter $a$, only $\Psi_k(\lambda)$ does.) The
cutoff $J_{e}$ is noise dependent and is given by

\begin{equation}
 \label{SignalToNoiseConditionExponential}
 \frac{1}{\alpha_{J_e}}<SNR<\frac{1}{\alpha_{(J_e+1)}}\;.
 \end{equation}

The cutoffs $J_p$ and $J_e$ are of prime importance because they
measure the number of degrees of freedom in the recovered
spectrum. Desirable is that both cutoffs be as close as possible
to the number of measurements, $N$, which is the case when the
Signal to Noise Ratio (SNR) is high. Although the equations
(\ref{SignalToNoiseConditionPeriodic2}) and
(\ref{SignalToNoiseConditionExponential}) look formally similar,
they give completely different solutions to the cutoffs. This is a
consequence of the different rate at which the numbers $\alpha_k$
and $\beta_k$ decrease to zero which we will study in the next
section.

\subsection{The number of replicates }

The SNR dictates how many spectral components are reliable and can
be use to recover the spectrum. We can imagine that by using
replicates we can improve the SNR and so the two cases will come
close to each other.This is not true; actually  we need an
unrealistic number of replicates to keep even few components for
the exponential case. Indeed, with the help of $r$ replicates, the
SNR increase to
\begin{equation}\label{Improved SNR}
   SNR\sqrt{r}\;,
\end{equation}
and the equations for the number of components $J_e,J_p$ to enter
into the recovered spectrum are
\begin{eqnarray}\label{ComponentEquation}
    \frac{1}{\alpha_{J_e}}\leq SNR
    \sqrt{r}<\frac{1}{\alpha_{J_e+1}}\;,\\
\frac{1}{\beta_{J_p}}\leq SNR
    \sqrt{r}<\frac{1}{\beta_{J_p+1}}\;.
\end{eqnarray}
The plots , Fig 7, of the number of replicates $r$ as a function
the number of spectral components $J_e$ or $J_p$ reveal that using
a periodic stimulus we can use many more spectral components to
recover the spectrum. The number of replicate growth very fast in
the exponential case (for $SNR=10$ we need 269 replicates for 4
spectral components), whereas in the periodic case, the number of
replicates stays low for many spectral components ( only for the
17th component it raises to 14, with $SNR=10$).

 The source of such a discrepancy
is that the eigenvalues $\alpha_{k}$ tend fast to zero as
\begin{equation}
\label{A} {\alpha_{{k}}}^{2}= {\frac {\pi }{\cosh \left( \pi
\,\xi_k \right) }}
\end{equation}
where $\xi_k$ tends to infinity like a polynomial of degree at
least one in $k$ (there is no analytical formula for $\xi_k$). For
the plotted example, $\gamma=5,q=1/20,\;\Delta=60^{1/20}$

\begin{eqnarray}
  \alpha_0 &=& 7.66\cdot10^{-1} \;,\\
  \alpha_1 &=& 3.28\cdot10^{-2} \;,\\
  \alpha_2 &=& 1.02\cdot10^{-3} \;,\\
  \alpha_3 &=& 1.74\cdot10^{-5} \;,\\
  \alpha_{19}&=& 9.94\cdot10^{-29}\;.
\end{eqnarray}

 For the periodic case the situation is much better. Here the numbers $\beta_k$ depends only
on the product $\tau \Omega$ and so is customary to introduce the
parameter $w$ through $2\pi w=\tau \Omega.$ Then, for $w=1/3$ for
example, we get

\begin{eqnarray}
  \beta_1 &=& 0.99 \;,\\
  \beta_2 &=& 0.99 \;,\\
  \beta_3 &=& 0.99 \;,\\
  \beta_4 &=& 0.99 \;,\\
  \beta_{20} &=& 0.000084\;,
\end{eqnarray}
There is no an general analytical formula for $\beta _k$ but it
was proven that the first $2Nw$ beta numbers are close to 1 with
the rest of them decreasing fast to zero. The fact that the
majority of the eigenvalues for the periodic case are 1 whereas
the eigenvalues for the exponential case decrease fast to zero is
the source of the difference between the two cases.
\begin{figure}[h]
\centering
\includegraphics[width=10cm]{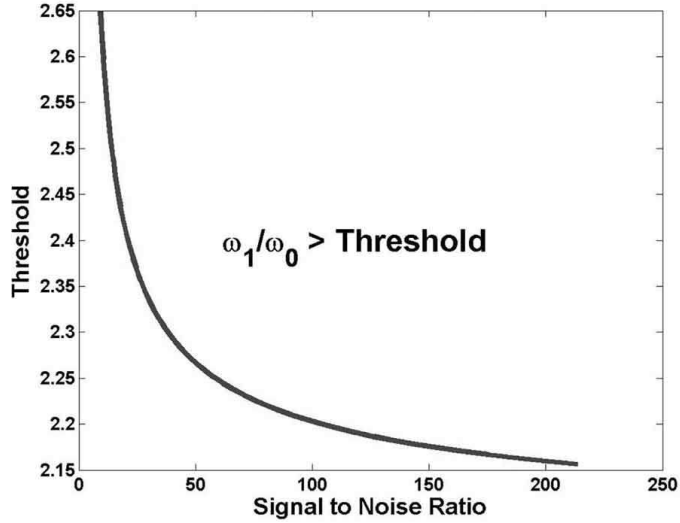}
\caption{The Threshold as a function of SNR}
\end{figure}

 Another interesting question is related to the resolution of
the different exponentially decaying signals  present in the
output signal. For the periodic case we do not address this
question, because the output signal has the same frequency as the
input periodic  signal (after the transients are gone). However,
for the response to a step stimulus, the transients contain the
information. To obtain this information we have to resolve the
transient components.The resolution power depends on the Signal to
noise Ratio (SNR). For example, to be able to resolve the decay
rates $\lambda_1$ and $\lambda_2$ when they are real positive
numbers, we need to have

\begin{equation}\label{Resolve1}
    \frac{\omega_1}{\omega_0}>Threshold(SNR)\;.
\end{equation}

The Threshold as a function of SNR is plotted in the figure.
Notice that we work with real $\lambda_{1,2}$ so $\omega_1 \geq
2\omega_0 $ for all $SNR.$
%\begin{figure}[!h]
    %\begin{center}
        %\includegraphics{C:/fig9999PNAS.pdf}
        %\includegraphics{C:/PNAS/Fig6.pdf}
    %\end{center}
    %\caption{The Threshold as a function of SNR.}
%\end{figure}

\vspace{1cm}

\appendix{\bf {Appendix 1. The eigenvalue problem for the periodic case}}

Recall that the measured quantities $r(n)$ for $n=0\dots
N-1,\;$can be expressed as

\begin{equation}
r \left( n \right) =\int
_{-\Omega}^{\Omega}\!{e^{in\tau\,\omega}}S_{per} \left( \omega
\right) {d\omega}\;.
\end{equation}
From the N data points we can find a N-dimensional approximation
to the spectrum $S_{per}(\omega)$ solving the following singular
value problem: find $V_k(\omega)$,$v_k(n)$ and $\lambda_k$ that
satisfy
\begin{eqnarray}
\label{SingularProblemPeriodic}
 {\cal{L}}V_k&=&\lambda_k v_k\;,
\\
{\cal L}^{*}v_k&=&\lambda_kV_k\;,
\end{eqnarray}
where the operator $\cal L$ and its conjugate $\cal L^*$ are
\begin{eqnarray}
({\cal L}f)(n)&=&\int _{-\Omega}^{\Omega}\!{e^{in\tau\,\omega}}f
\left( \omega \right) {d\omega}\;,
\\
({\cal L^*}g)(\omega)&=&\sum _{n=0}^{N-1}{e^{-in\tau\,\omega}}g
\left( n \right)\;.
\end{eqnarray}
The set $V_k(\omega)$ form an orthonormal basis in
$L^2(-\Omega,\Omega)$ and $v_k(n)$ an orthonormal basis in the
euclidian space $E^N.$ In the $V_k$ basis, the N-dimensional
approximation to the spectrum reads like
\begin{equation}
\label{SolutionSpectrum}
 S_{per}^{N} \left( \omega \right) =\sum
_{k=0}^{N-1}{\frac {r_{{k}}}{ \lambda_{{k}}}}V_k(\omega)\;,
\end{equation}
where the coefficients $r_k$ are obtained from the decomposition
of the measured data $r(n)$

\begin{equation}
r \left( n \right) =\sum _{k=0}^{N-1}r_{{k}}v_{{k}} \left( n
\right)\;.
\end{equation}
The solution to the singular problem
(\ref{SingularProblemPeriodic}) can be reduced to the eigenvalue
problem for the operator ${\cal LL^*}$: find the eigenfunctions
$v_k(n)$ and the eigenvectors $\lambda_k^2$ from

\begin{eqnarray}
\label{slepian}
 \sum _{m=0}^{N-1}{\frac {\sin \left(\tau \,\Omega
\left( m-n \right)
 \right) }{\pi \, \left( m-n \right) }}v_k(m)=\beta_k^2 v_{{k}} \left( n
 \right)\;,
\end{eqnarray}

where

\begin{equation}
\beta _k=\sqrt {\frac {\tau}{2\pi}}\lambda_k\;.
\end{equation}

 In this way, the solution to our problem is reduced to the
diagonalization of an $N\times N$ matrix. This is the famous
problem, \cite{Slepian}. We have two independent parameters $\tau
$ and $\Omega.$ The eigenvalues of the problem ($\ref{slepian}$)
depends on $w$, defined as $2\pi w=\tau \Omega .$  The first $2Nw$
eigenvalues are close to $1$ with the rest of them close to zero.
As a consequence, from ($\ref{SolutionSpectrum}$) we see that we
can keep only the first $2Nw$ terms, because the rest of them are
highly amplified by the small values of the eigenvalues which is
dramatic when the values $r_k$ are corrupted by noise. We want
than $2Nw$ to be close to $N$ which case $w=1/2$ and
$\Omega=\pi/\tau.$ This situation corresponds to a sampling
parameter $\tau$ tuned for recovering the spectrum up to the
frequency $\Omega $. In general case when $\Omega \geq \pi/\tau$.
The recovered spectrum, when noise is present will be than

\begin{equation}
\label{SolutionSpectrumWithError}
 {\tilde
S}_{per}^{N}\left( \omega \right) =\sum _{k=0}^{N-1}{\frac
{r_{{k}}+\epsilon_k}{ \lambda_{{k}}}}V_k(\omega).
\end{equation}

To connect with the notations from Section 10, denote
$\Phi_k(\omega)=\sqrt {\tau/2\pi} V_k(\omega )$ and $s_k =
{r_k/\beta _k}.$

\appendix{\bf {Appendix 2. The eigenvalue problem for the step
stimulus}}

The problem for the exponential decay responses was solved in
\cite{Italy}. The unknown spectrum $S_{exp}$ and the measured data
$r(n)$ are connected through the equation

\begin{equation}
r \left( n \right) =\int _{1}^{\gamma}\!{e^{-ap_{{n}}x}}S_{exp}
\left( ax \right) a{dx}\;,
\end{equation}

with $\gamma=b/a$ and $p_n=(q/a)\Delta^n, n=1,\dots,N.$  Like for
the periodic case, an N-dimensional approximation to the spectrum
can be found from the solutions of two coupled equations:
\begin{eqnarray}
{\cal K}U_{{k}}=\alpha_{{k}}u_{{k}}\;,
\\
{\cal K^{*}}u_{k}=\alpha_{k}U_{{k}}\;,
\end{eqnarray}

where
\begin{eqnarray}
({\cal K}f)(n)=\int _{1}^{\gamma}\!{e^{-ap_{{n}}x}}f \left( x
\right) {dx}\;,
\\
({\cal K^{*}}g)(x)=\sum _{n=1}^{N}w_{{n}}g \left( {{n}} \right)
{e^{-ap_{{n}}x}}\;,
\end{eqnarray}
with the weights given by $w_n=p_n\ln(\Delta )$, see
$\cite{Italy}.$ The unknowns are the functions  $U_{k}(x)$ that
form an orthonormal basis in $L^2(1,\gamma)$ and the functions
$u_{k}(n)$ that form a basis in the euclidian space $\mathbb{R}^N$
endowed with the scalar product
\begin{equation}
(g,h)=\sum _{n=1}^{N}w_{{n}}g \left( {{n}} \right) h \left( {{n}}
\right)\;.
\end{equation}
The N-dimensional approximation to the unknown spectrum can now be
written as a decomposition in $U_{k}$ basis as

\begin{equation}
\label{SolExp} S_{exp}^N \left( \lambda \right) =\frac{1}{a}\sum
_{k=1}^{N}{\frac {r_{{k}}}{\alpha_{{k }}}}  U_{{k}}   \left(
{\lambda/a} \right)\; ,
\end{equation}

with the components $r_k$ obtained from decomposing the measured
data $r_n$ in the basis $u_{k}$
\begin{equation}
r_{{k}}=\sum _{n=1}^{N}w_{{n}}r \left( n \right)   u_{{k}}
   \left( n \right)\; .
\end{equation}
Similar to the periodic case, the eigenvalue problem to be solved
now is

\begin{eqnarray}
\label{EigenvalueExponential}
 \sum _{m=1}^{N}{\frac {\sqrt {w_n  w_ m }}{a}}\;{\frac {{e^{-a \left( p_ n  +p_ m
\right) }}-{e^{-b \left( p_ n +p_ m \right) }}}{p_ n  +p_ m
}}\bar{u}_{k}(m)=\alpha_{k}^2\bar{u}_{k}(n)
\end{eqnarray}
with $\bar{u}_{k}(n)=\sqrt{w_n}\,u_{k}(n).$ The matrix that is
diagonalized in the problem (\ref{EigenvalueExponential}) is a
symmetrized version of  $\cal {KK^*}$ and so there is a scaling
difference between $u_k$ and ${\bar u}_k$.
 The eigenvalues $\alpha_{k}$ tend fast to zero as
\begin{equation}
\label{A} {\alpha_{{k}}}^{2}= {\frac {\pi }{\cosh \left( \pi
\,\xi_k \right) }}\;,
\end{equation}
where $\xi_k$ tends to infinity like a polynomial of degree at
least one.
 The recovered spectrum is
\begin{equation}
\label{SolExpError} {\tilde S}_{exp}^N \left( \lambda \right)
=\frac{1}{a}\sum _{k=1}^{N}{\frac {r_{{k}}+\epsilon_k}{\alpha_{{k
}}}} U_{{k}}   \left( {\lambda/a} \right)\;,
\end{equation}
where the terms $\epsilon_k$ are due to random experimental
errors.

In Section 10 we write the spectrum in terms of
$\Psi_k(\lambda)=(1/a)U_k(\lambda/a)$ and $s_k=r_k/\alpha_k.$

\appendix{\bf {Appendix 3. The eigenvalue problem for continuous measurements}}

 We discussed the spectrum recovery from a finite number of data, which is the case of laboratory measurements. However, it is instructive to
inspect the case when we know $f(t)$ from ($\ref{generalproblem}$)
for all $t$ and in the limit for which $a=0$, $b=\infty $ and
$\Omega=\infty$. This problem was studied in \cite{Pike}. As a
bonus, we get an expression for the resolution of the exponential
spectrum and a direct understanding of the difference in the two
eigenvalue problems presented in Appendix 3 and 4.
 The solution for the exponential case is
 in terms of the eigenvalues and eigenfunctions of the
kernel $K(\mu t)$
\begin{equation}
\int _{0}^{\infty}\!K \left( \eta\,t \right) \Phi_{{n}} \left(
\eta \right) { d\eta}=\Xi_{{n}}\Phi_{{n}} \left( t \right)\;.
\end{equation}
The eigenfunctions form an orthogonal basis and both the measured
data $f_{exp}(t)$ and the unknown $S_{exp}(\mu)$ can be decomposed
as:
\begin{eqnarray}
f_{exp} \left( t \right) &=&\sum _{k=1}^{\infty
}f_{{exp,k}}\Phi_{{k}} \left( t \right)\;,
\\
S_{exp} \left( \eta \right) &=&\sum _{k=1}^{\infty
}S_{{exp,k}}\Phi_{{k}} \left( \eta \right)\;.
\end{eqnarray}
Now we have the relation between the spectrum and the measured
data
\begin{equation}
S \left( \eta \right) =\sum _{k=1}^{\infty }{\frac
{f_{{exp,k}}}{\Xi_{{k}}} }\Phi_{{k}} \left( \eta \right)\;,
\end{equation}
with $\Xi_k$ arranged in decreasing order $\Xi_1>\Xi_2>\dots \,.$
We see from this expression that if $\Xi_k$ decrease to zero and
the components of the measured data $f_k$ are corrupted by noise,
than the components with large $k$ cannot be used to recover
$S_{exp}(\eta)$. The function thus recovered $S_{exp}(\eta)$ has
information just from the first components $f_{exp,k}$. Only if
the eigenvalues don't decrease to zero we can use all the terms in
the decomposition.

For the exponential decay problem (\ref{int}) the eigenvalues form
a continuous spectrum ( k is a positive real number)
\begin{equation}
  \left|\, \Xi _k   \,\right| \, ^{2}={\frac
{\pi }{\cosh \left( \pi \,k \right) }}
\end{equation}
For the periodic solution (\ref{intp}) with $\Omega =\infty $
(Fourier transform) the spectrum is discrete ( $k=0,1,\dots,\infty
$)
\begin{equation}
\Xi_k=-i^k\sqrt{2\pi}
\end{equation}

It is obvious the difference between the exponential decay
situation ( step stimulus) and the periodic response. In the
former case the eigenvalues tend fast to zero whereas in the later
case they never approach zero ( they have a constant modulus one.)

We aim now to find the resolution limit for resolving the
exponential decay problem \cite{Italy}. For a given signal to
noise ration $SNR$ we want to find the minimum ratio of the
exponential decay rates $\eta_i/\eta_{i+1}$ that can be resolved.
 Going a little deeper into the solution of the exponential decay
 response,\cite{Pike}
 we find that the decomposition of the spectrum $g(\mu)$ is
 \begin{equation}
S_{exp} \left( \eta \right) =\int _{0}^{\infty
}\!{a_{{k}}}\!^{+}\,{\Xi_{{k}}}\!^{+ }\,\Phi_{{k}}\!^{+} \left(
\eta \right) {dk}+\int _{0}^{\infty }\!{a_{{k}}}\!^{-}\,
{\Xi_{{k}}}\!^{-}\,\Phi_{{k}}\!^{-} \left( \eta \right) {dk}
 \end{equation}
where the eigenfunctions are
\begin{eqnarray}
{\Phi_{{k}}}\!^{+}(\eta)&=&\,\frac{1}{\sqrt{k\,\pi}}\cos \left(
k\ln \left( \eta \right) -\frac{\theta}{2}\,
 \right)
 \\
 {\Phi_{{k}}}\!^{-}(\eta)&=&\!\!-\frac{1}{\sqrt{k\,\pi}}\sin \left( k\ln
\left( \eta \right) -\frac{\theta}{2}\,
 \right)
\end{eqnarray}
with the angle $\theta$ expressed in terms of the Gamma function
\begin{equation}
\theta={\it angle} \left( \Gamma  \left( 1/2+ik \right) \right)\,.
\end{equation}
 Due to noise, we can recover the components
up to a maximum $k_0$, so we have all the components with $k<k_0.$
For this reason, we only can resolve points on the axis $\eta$
that are separated at a distance larger than the distance between
two zeros of $\Phi_{k_0}.$ Due to the presence of $ln(\eta)$ in
the argumet of the trigonometric function, the zeros are
\begin{equation}
\mu_{{m}}={e^{{\frac {1/2\,\theta+m\pi }{k_0}}}}
\end{equation}
To conclude, two decay rates $\eta_a$ and $\eta_b$ can be
recovered from the measured data if
\begin{equation}
\label{resolvability}
 \frac{\eta_a}{\eta_b}>{\frac
{\eta_{{m}}}{\eta_{{m+1}}}}={e^{\left({{\frac {\pi }{
k_{{0}}}}}\right)}}
\end{equation}
The value $k_0$ that is the index for the maximum eigenvalue
recoverable from noise is given by comparing the signal to noise
ratio with the eigenvalue
\begin{equation}
{\frac {\cosh \left( \pi k_{{0}} \right) }{\pi }}={{\it SNR}}^{2}
\end{equation}
Applying (\ref{resolvability}) to the example we work with
(\ref{lambda1lambda2}) we obtain the condition
\begin{equation}
{\frac {\omega_{{1}}}{\omega_{{0}}}}>2\,\cosh \left( {\frac {\pi
}{2\,k_{{0}}}} \right)
\end{equation}

\end{document}